\documentclass[useAMS,usenatbib]{mn2e}
\usepackage{graphicx,amsmath,multirow,amssymb}
\usepackage{subfig}
\usepackage{natbib}
\newcommand{\comment}[1]{}

\newcommand\Msun{M_{\odot}}
\def\simgt{\lower.5ex\hbox{$\; \buildrel > \over \F sim \;$}}
\def\simlt{\lower.5ex\hbox{$\; \buildrel < \over \sim \;$}}

\title[Obscured C-stars in the MCs]{On the nature of the most obscured C-rich AGB stars in the Magellanic Clouds}
\author[Ventura et al.]{P. Ventura$^1$, A. I. Karakas$^{2,7}$, F. Dell'Agli$^{1}$, D. A. Garc\'{\i}a--Hern\'andez$^{3,4}$,  
\newauthor
M.~L. Boyer$^{5,6}$, M. Di Criscienzo$^1$ \\
$^1$INAF -- Osservatorio Astronomico di Roma, Via Frascati 33, 00040, Monte Porzio Catone (RM), Italy \\
$^2$Research School of Astronomy and Astrophysics, the Australian National University, Canberra, ACT 2611, Australia \\
$^{3}$Instituto de Astrof\'{\i}sica de Canarias, E-38200 La Laguna, Tenerife, Spain \\
$^{4}$Departamento de Astrof\'{\i}sica, Universidad de La Laguna (ULL), E-38206 La Laguna, Tenerife, Spain \\
$^5$Observational Cosmology Lab, Code 665, NASA Goddard Space Flight Center, Greenbelt, MD 20771, USA \\
$^6$CRESST and Department of Astronomy, University of Maryland, College Park, MD 20742 USA  \\
$^7$Monash Centre for Astrophysics, School of Physics and Astronomy, Monash University, VIC 3800, Australia \\
}

\begin{document}

\date{Accepted, Received; in original form }

\pagerange{\pageref{firstpage}--\pageref{lastpage}} \pubyear{2012}

\maketitle

\label{firstpage}

\begin{abstract}
The stars in the Magellanic Clouds with the largest degree of obscuration
are used to probe the highly uncertain physics of stars in the asymptotic giant
branch (AGB) phase of evolution. Carbon stars in particular, provide key information
on the amount of third dredge-up (TDU) and mass loss.
We use two independent stellar evolution codes to test how a different treatment
of the physics affects the evolution on the AGB. The output from the two codes are
used to determine the rates of dust formation in the circumstellar envelope, where the
method used to determine the dust is the same for each case.
The stars with the largest degree of obscuration in the LMC and SMC are identified as
the progeny of objects of initial mass $2.5-3~M_{\odot}$ and $\sim 1.5~M_{\odot}$,
respectively. This difference in mass is motivated by the difference in the star
formation histories of the two galaxies, and offers a simple explanation of the redder
infrared colours of C-stars in the LMC compared to their counterparts in the SMC.
The comparison with the Spitzer colours of C-rich AGB stars in the SMC shows that a minimum
surface carbon mass fraction $X(C) \sim 5\times 10^{-3}$ must have been reached by
stars of initial mass around $1.5~M_{\odot}$.
Our results confirm the necessity of adopting low-temperature opacities in stellar evolutionary
models of AGB stars. These opacities allow the stars to obtain mass-loss
rates high enough ($\gtrsim 10^{-4}M_{\odot}/yr$) to produce the amount of dust needed
to reproduce the Spitzer colours.
\end{abstract}

\begin{keywords}
Stars: abundances -- Stars: AGB and post-AGB, galaxies: Magellanic Clouds 
\end{keywords}

\section{Introduction}
The recent years have witnessed a growing interest in the evolution of stars of low
and intermediate mass ($M < 8~M_{\odot}$), which evolve through the asymptotic giant branch (AGB)
after the end of core helium burning.  While this evolutionary phase only accounts for a
very small percentage of the overall lifetime, it is extremely important
because it is during the AGB when the richest nucleosynthesis occurs as well as the most intense
mass loss. This means that AGB stars are able to enrich their environment with gas 
chemically altered by internal nuclear processes and with the dust formed in their wind.

The unsolved problems of galaxy formation and chemical evolution at all redshifts requires
a full understanding of the topic of stellar evolution and nucleosynthesis of AGB stars.
Particular areas where AGB stars have proven useful include inferring the masses of galaxies
at high redshifts, owing to their large infrared luminosities (Maraston et al. 2006);
chemical evolution of galaxies (Romano et al. 2010), owing to the ability of AGB stars to
enrich their environment with stellar winds; 
because of the efficiency of the dust formation process in their winds, AGB stars play a 
crucial role in the formation and evolution of galaxies (Santini et al. 2014); 
recent studies suggested a relevant contribution from AGB stars to the dust 
content of high-redshift quasars (Valiante et al. 2011),
contrary to earlier investigations, which stressed the dominant role of supernovae (SNe)
\citep{todini01, nozawa03, maiolino04}; finally, AGB 
stars $\gtrsim 5~M_{\odot}$ are one of the favoured polluters for providing the
gas required to form second generation stars in globular clusters \citep{ventura01}.

While theoretical models of AGB stars have seen significant progress over the last few years
(e.g., Karakas \& Lattanzio 2014), the results are still not completely reliable. This is
primarily because we still do not understand how mass loss or convection work in stars.
The problem with convection manifests itself in two important ways in AGB stars. The first 
is through the efficiency of third dredge-up (TDU), which is the inward penetration of the
convective envelope following each thermal pulse. The efficiency or depth of TDU depends on how 
convective borders are treated numerically \citep[e.g.,][]{frost96}.
The second is through temperature gradients in convective regions, which cannot be 
calculated from first principles (Ventura \& D'Antona 2005).  
Therefore comparing theoretical AGB models with observations is crucial in order to 
substantially improve the predictive power of the stellar evolution models.

The Magellanic Clouds (MCs) are an ideal environment to test theoretical predictions, 
because they are relatively close (51 kpc and 61 kpc respectively, 
for the LMC and SMC, Cioni et al. 2000; Keller \& Wood 2006) and the low reddening 
($E_{B-V}=0.15$~mag and 0.04 mag, respectively, for the LMC and SMC, Westerlund
1997). Furthermore, the study of the MCs provides a wider range of possibilities in
comparison with the Milky Way, because the interstellar medium of our Galaxy is highly 
obscuring and the distances of the stars are unknown. Also, the MCs probe a lower 
metallicity than the disc of our Galaxy.

Various dedicated surveys have been devoted to observe the AGB population of the MCs:
the Magellanic Clouds Photometric Survey \citep[MCPS,][]{zaritsky04}, the Two Micron All 
Sky Survey \citep[2MASS,][]{skrutskie06}, the Deep Near Infrared Survey of the Southern 
Sky \citep[DENIS,][]{epchtein94}, the Surveying the Agents of a Galaxy's Evolution 
Surveys with the {\it Spitzer telescope} for the LMC \citep[SAGE--LMC,][]{meixner06}
and the SMC \citep[SAGE--SMC,][]{gordon11}, and {\it HERschel} 
Inventory of The Agents of Galaxy Evolution \citep[HERITAGE,][]{meixner10, meixner13}.

These surveys have produced a wealth of data on AGB stars, which can be used to compare
to the results of theoretical calculations. To reproduce the observed infrared (IR) 
colours, which are sensitive to the amount of dust formed in the circumstellar envelope, 
some research groups couple the description of the internal star with the dust formation 
process in the wind. This approach was first set up by the Heidelberg 
team \citep{fg01, fg02, fg06} and later used by other independent 
investigators \citep{paperI, paperII, paperIII, paperIV, nanni13a, nanni13b, nanni14}.

Against this background, we have started a research project aimed at constraining 
the uncertain physics of AGB evolution, by focusing our attention on the most obscured 
AGB stars in the MCs.

This study follows \citet{marigo08}, who built theoretical isochrones which included
the AGB phase and extended the isochrones to the IR bands.  Here we take a step forward by
relaxing some of the assumptions made by \citet{marigo08}. These include determining the expansion
velocity of the wind and the final dust to gas ratio, and we model the AGB phase by 
the full integration of the stellar structure equations, rather than using a synthetic 
approach. We base our analysis on AGB models calculated from two independent stellar
evolution codes which made a significant contribution to the literature of AGB stars over the
past couple of decades:  MONASH \citep{frost96} and ATON \citep{ventura98}.
The two codes have been developed and updated independently and include 
numerous differences between each other in the numerical structure and in the physical 
ingredients used. This procedure provides us with a much more complete and critical 
analysis because: a) it adds robustness to common findings; b) the discrepancy between 
results provide an estimate of the uncertainty associated with that result; and c) the 
comparison with the observations may allow us to select the most appropriate description 
of a given phenomenon (e.g., the treatment of convective borders).
 
In the first paper of this series \citep{ventura15a} we focused on oxygen-rich stars 
in the LMC with the brightest IR emission, interpreted as the progeny of $5-6~M_{\odot}$
stars experiencing hot bottom burning (HBB). 

In this paper we turn to the carbon star population in the MCs. 
The goal of the present investigation is to use the sample of numerous C-rich AGB stars 
to reconstruct the various stages of their evolutionary history. In particular, we are
interested in studying the amount of carbon accumulated in the external mantle 
and the rate at which their envelope has been gradually lost by stellar winds. 

We will preferentially concentrate on the C-stars with the largest degree of obscuration, 
for which the mass and chemical composition of the progenitors can be established; 
this makes the comparison with the models easier and more reliable.

The paper is organised as follows. Section \ref{mcs} presents an overview of the
observations of the AGB population of the MCs, with a summary of the interpretative
analysis proposed so far. The physical ingredients used to
calculate the evolutionary sequences presented here and the method followed to
describe the dust formation process and to produce synthetic spectra are given 
in Section \ref{inputs}. Section \ref{evolution} presents the results concerning
the main evolution and dust properties of the stars currently evolving through the
C-rich AGB phase in the MCs. The comparison between the AGB stars with the largest degree
of obscuration and the results from theoretical models is given in Section \ref{colours},
while Section 6 discusses our interpretation in terms of other observational parameters as 
well as future observations to test it. Conclusions are presented in Section \ref{concl}.

\section{AGB stars in the Magellanic Clouds}
\label{mcs}
The results from the surveys MCPS, 2MASS and DENIS have been used to derive information on the 
internal structure and on the efficiency of the mechanisms that are altering the surface 
chemical composition of AGB stars.

Studies focused on the interpretation of the luminosity function of carbon stars in the 
LMC and SMC have provided important information on the efficiency of 
TDU and the core mass at which TDU begins \citep{martin93,marigo99, karakas02, izzard04, 
marigo07, stancliffe05}.

The mid-IR {\it Spitzer} data, combined with near-IR 2MASS photometry, have been extensively used to
derive the main properties of the stars observed, specifically the luminosity, the rate
of mass loss, and the dust injection rate
\citep{martin07, riebel10, riebel12, srinivasan09, srinivasan11, boyer11, boyer12}.
Various methods have been proposed to classify the observed stars and
separate carbon stars from their oxygen-rich
counterparts \citep{cioni00a, cioni00b, cioni06}. The stars with the highest
degree of obscuration have been traditionally referred to as "extreme" \citep{blum06}, 
owing to their extremely red IR colours and the uncertain surface chemical composition.

In a series of papers \citep{flavia14, flavia15a, flavia15b} we tackled the
problem of interpreting the IR observations of the MCs AGB star population by 
confronting the observational evidence with results of full AGB evolutionary models.
The models also account for dust formation in the circumstellar envelope. 
The latter ingredient is mandatory because radiation from the central object
is reprocessed to longer, infrared wavelengths by dust.

Based on the results by \citet{flavia15a}, \citet{ventura15a} performed an extensive 
exploration of the various input used to build AGB evolutionary sequences of oxygen-rich 
intermediate-mass stars in the LMC. The study, which used results from different evolutionary
codes, found that the expected position in the various IR colour--colour 
planes occupied by oxygen-rich AGB stars during the phase with the strongest 
obscuration is practically independent of the details of AGB modelling.

In this paper we turn our attention to the C-star population in the LMC and SMC.
\citet{flavia14} characterised the AGB stars 
populating the diagonal band in the CCD of the two galaxies, extending to 
$[3.6]-[4.5] \sim 3$ \citep[LMC,][]{srinivasan11} and $[3.6]-[4.5] \sim 1.2$ 
\citep[SMC,][]{boyer11}, as carbon stars \citep[see also][]{flavia15a, flavia15b}. 
The scenario proposed by \citet{flavia14} is that the observed bands represent obscuration 
sequences: the stars move across the diagonal band as they become more enriched  
in carbon, owing to the effects of TDU. 
The subsequent investigations by \citet{flavia15a,flavia15b} followed a population synthesis
approach, which allowed us to determine the expected distribution of stars in the colour--colour 
and colour--magnitude planes obtained with the {\it Spitzer} bands. These studies are based
on the star formation history of the MCs by \citet{harris04, harris09} and on the evolutionary
times of stars of various mass and metallicity, reported in Table 1 in \citet{flavia15a}.

One general result is that the regions of the observational planes where highly-obscured
carbon stars are found are not well populated.
This is because the evolutionary time scales get shorter as the star becomes enriched in
carbon, owing to the considerable increase in the rate of mass loss. This has important
consequences on the (current) mass distribution of the carbon stars observed. All the
stars with initial mass in the range $\sim 1.25-3~M_{\odot}$ evolve as carbon stars from
a given stage of the evolution until the removal of the envelope, where the final
stellar mass is $\sim 0.6~M_{\odot}$.
On the other hand, for the reasons given above, the very 
final evolutionary phases have a very short duration. The combination of these two
factors makes the mass distribution of the LMC to peak around $\sim 1.3-1.4~M_{\odot}$, 
in nice agreement with the analysis based on pulsation periods by \citet{boyer15}.

\citet{flavia15a} found that in the LMC the C-rich AGB stars with the reddest
infrared colours formed during 
the $\sim 300$ Myr long epoch of high star formation, which occurred $\sim 500$ Myr 
ago, when most of the stars formed from gas with metallicity $Z=0.008$.
An age of $\sim 300-500$ Myr corresponds to stars with initial masses $\sim 2.5-3~M_{\odot}$.

By applying to the SMC the same population synthesis analysis, \citet{flavia15b} found
that the stars exhibiting the largest obscuration descend from $\sim 1.5~M_{\odot}$ stars,
formed $\sim 2$ Gyr ago, when the dominant metallicity of the interstellar medium was 
$Z=0.004$. Contrary to the LMC, only a negligible fraction of stars descending from 
$\sim 2.5-3~M_{\odot}$ objects are expected to evolve nowadays through the very final
AGB phases of highest obscuration, because the SFR of the SMC exhibits a much narrower
peak than the LMC at $\sim 500$ Myr ago.

We will focus on these masses and metallicities in the following sections.

\section{Numerical and physical inputs}
\label{inputs}
\subsection{Stellar evolution models}
This work is based on evolutionary sequences calculated with the ATON code \citep{ventura98} 
and with the Monash version of the Mount Stromlo Stellar Structure Program 
\citep[MONASH,][]{frost96}. 
Exhaustive discussions on the numerical and physical input adopted are found in the papers by 
\citet{ventura13} and \citet{karakas10}, which also provide a detailed description of the chemical 
and physical properties of the AGB evolution of these stars. 

The main differences between the two sets of models 
are the treatment of convection, the description of mass loss, and the calculation of
molecular opacities in the low-temperature region of the envelope:

\begin{itemize} 

\item{In the ATON code the convective instability is described by means of the Full
Spectrum of Turbulence model developed by \citet{cm91}, whereas in the MONASH case 
the traditional mixing length theory is used.}

\item{The ATON sequences are calculated by coupling nuclear burning and mixing of chemicals
in a diffusive-like scheme. Overshoot of convective eddies into radiatively stable regions
is described by means of an exponential decay of velocities from the convective/radiative
interface, fixed by the Schwarzschild criterion. The e-folding distance is assumed to
be $0.002H_p$ (where $H_p$ is the pressure scale height calculated at the formal boundary
of convection), in agreement with the calibration based on the observed luminosity
function of carbon stars in the LMC, given by \citet{paperIV}. The MONASH models 
of \citet{karakas10} 
assume instantaneous mixing and no overshoot beyond the formal Schwarzschild boundary
is applied. However, the MONASH models  implement an algorithm to search for a 
neutrally stable point \citep[e.g.][]{lattanzio86}, which has been shown to increase
the amount of TDU relative to models without \citep{frost96}. Note that in low-mass AGB
models the amount of TDU found by this algorithm is still less than required to match the 
C/O transition luminosity of MC AGB stars and some formal overshoot had to be applied 
\citep{kamath12}. 
}

\item{The mass loss rate for oxygen-rich models in the ATON case is determined via the 
\citet{blocker95} treatment; for carbon stars ATON uses the results from the Berlin group
\citep{wachter02, wachter08}. MONASH models rely on the classic description by \citet{VW93}.}

\item{In the ATON sequences, molecular opacities in the low-temperature regime (below 
$10^4$ K) are calculated by means of the AESOPUS tool \citep{marigo09}. 
The opacities are suitably constructed to follow the changes in the chemical composition 
of the envelope, particularly of the individual abundances of carbon, nitrogen, and oxygen.
The MONASH models from \citet{karakas10} include an approximate treatment for the molecular 
opacities (in particular CN, CO, H$_2$O and TiO) using the formulations from \citet{bessel89} with
the corrections by \citet{chiosi93}. These fits do include some compositional dependence, 
but do not account for large variations in the CNO species.} 

\end{itemize}

\begin{figure*}
\begin{minipage}{0.48\textwidth}
\resizebox{1.\hsize}{!}{\includegraphics{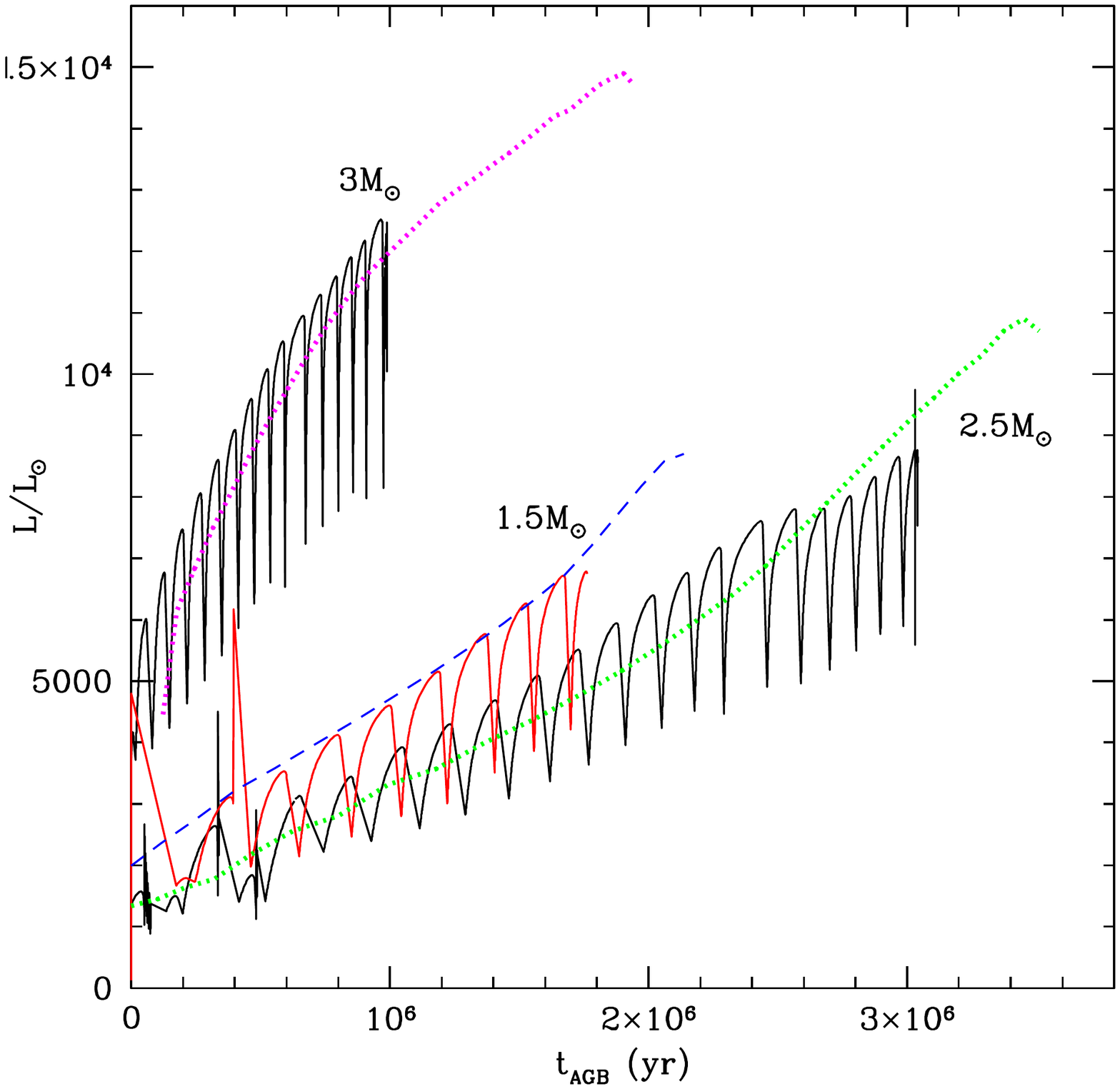}}
\end{minipage}
\begin{minipage}{0.48\textwidth}
\resizebox{1.\hsize}{!}{\includegraphics{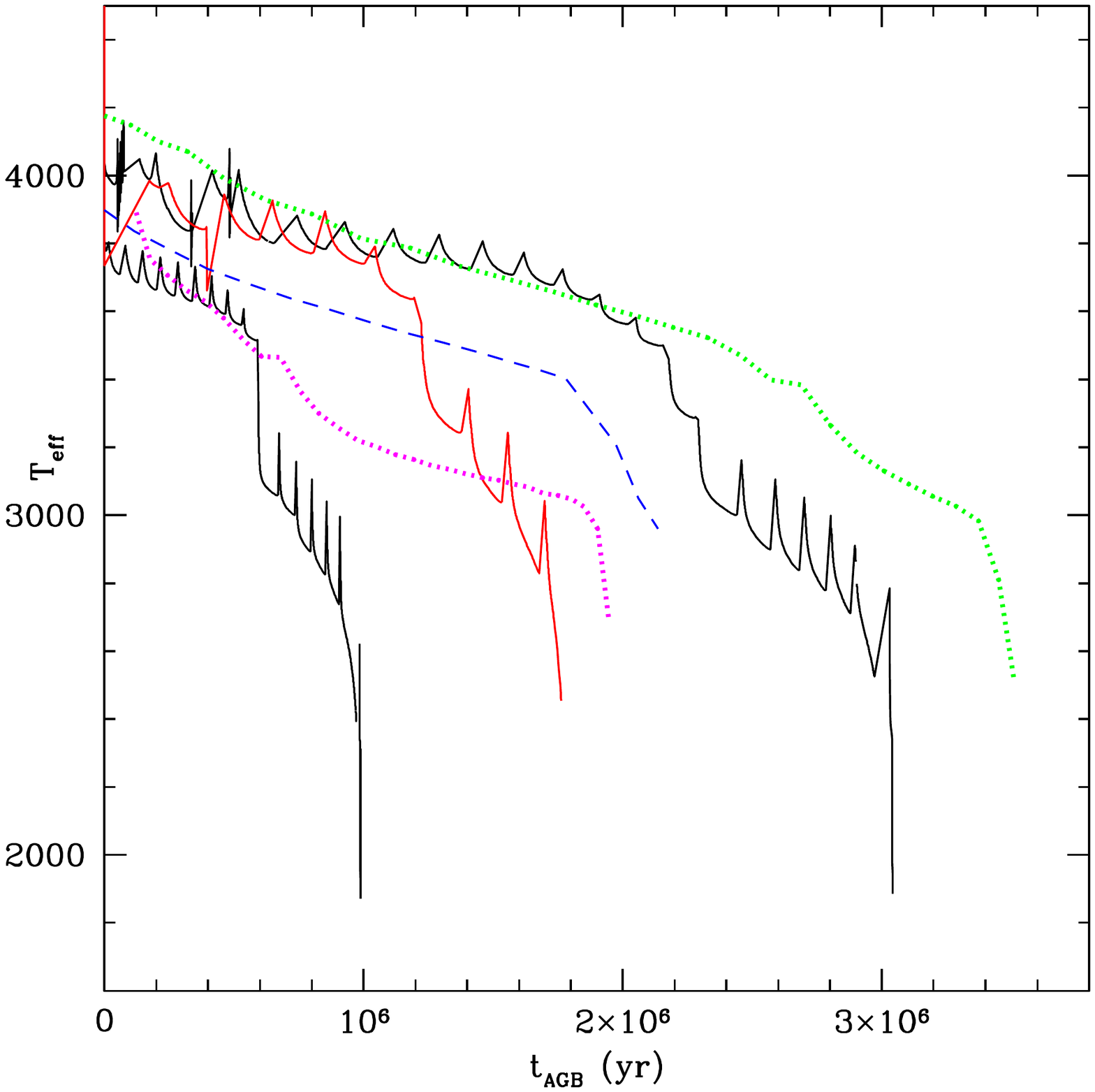}}
\end{minipage}
\vskip-70pt
\begin{minipage}{0.48\textwidth}
\resizebox{1.\hsize}{!}{\includegraphics{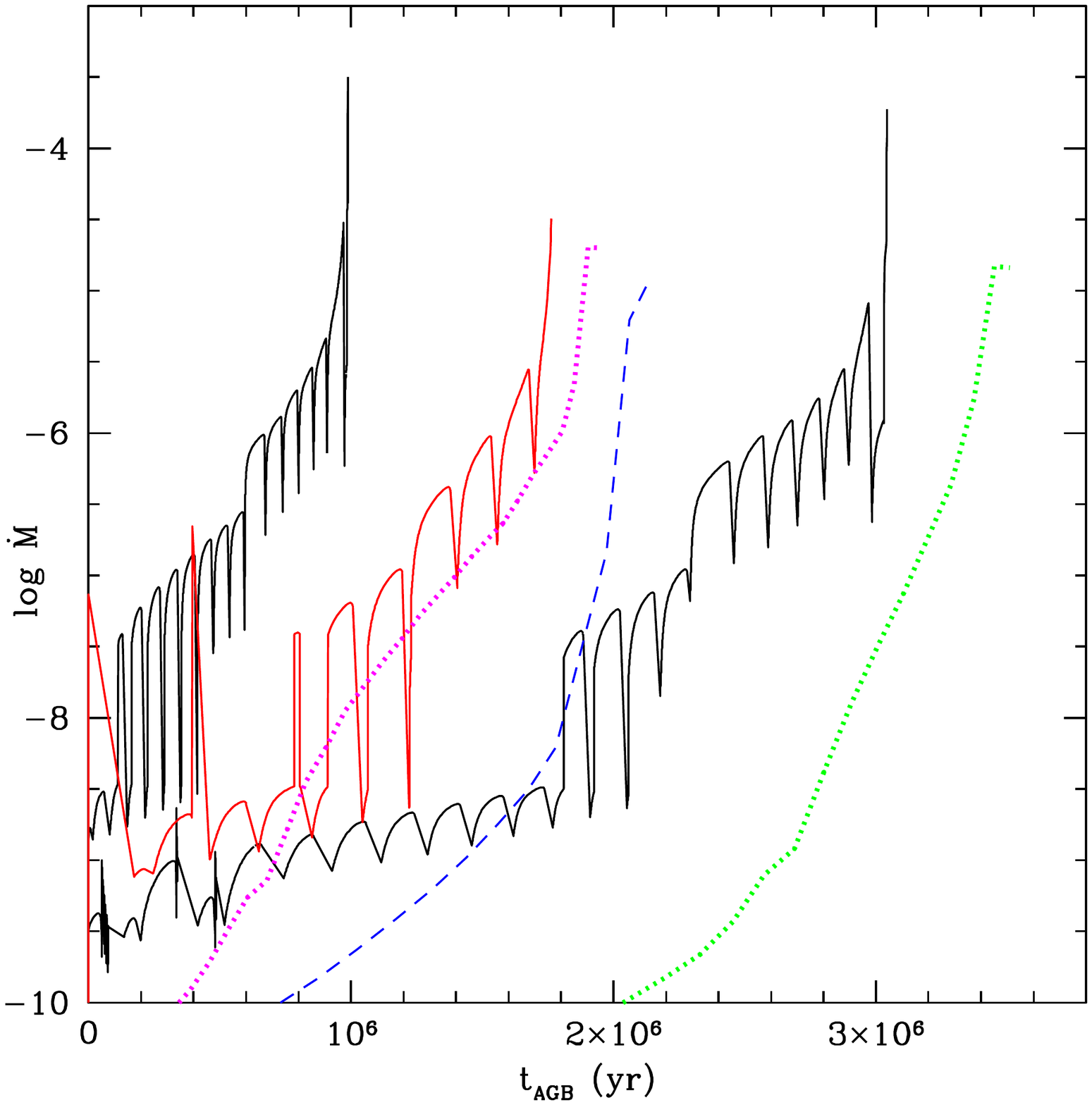}}
\end{minipage}
\begin{minipage}{0.48\textwidth}
\resizebox{1.\hsize}{!}{\includegraphics{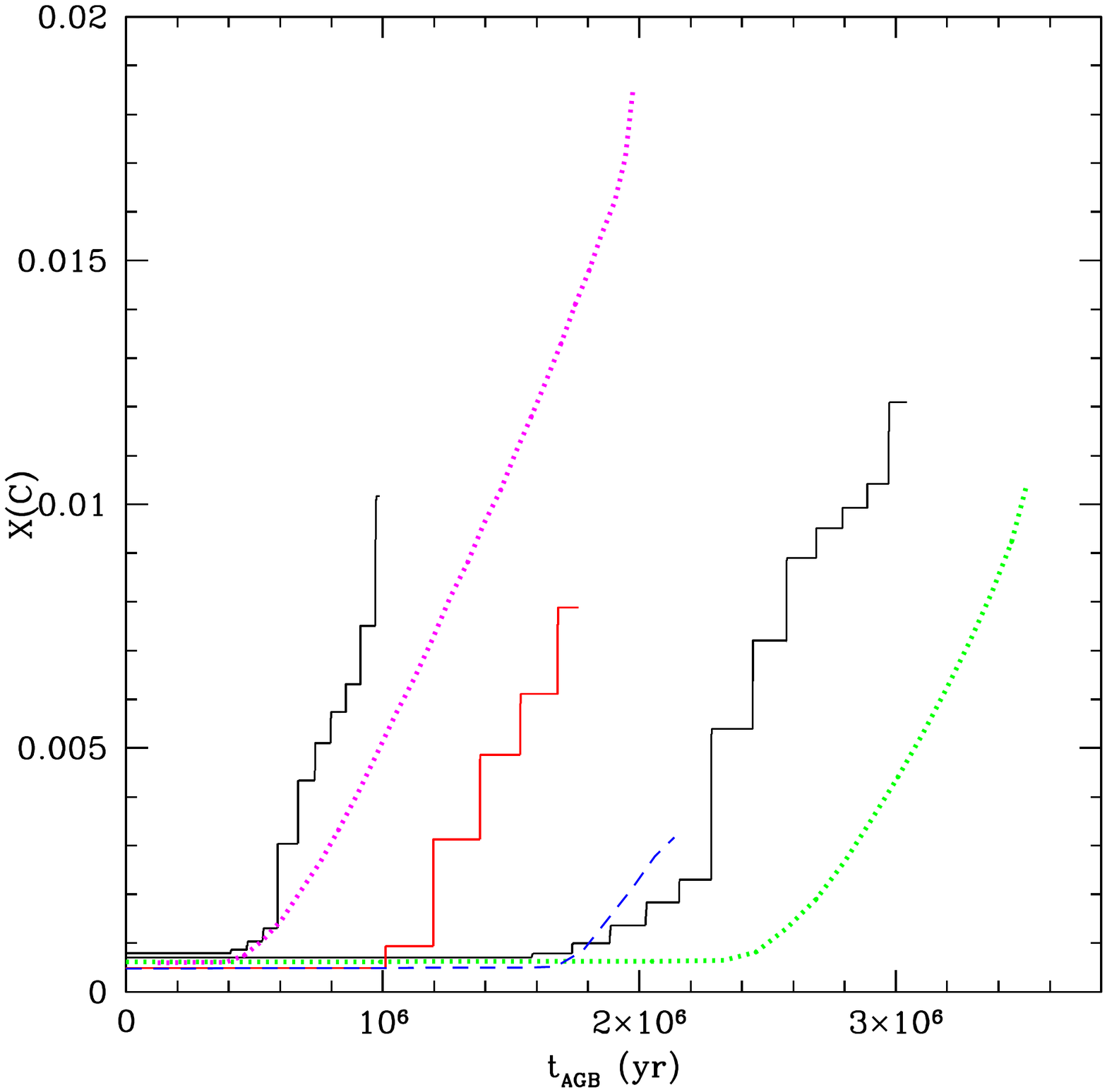}}
\end{minipage}
\vskip-50pt
\caption{The variation of luminosity (top--left panel), effective temperature (top--right),
rate of mass loss (bottom--left) and surface carbon mass fraction (bottom--right) 
during the AGB phase of models of various mass and metallicity, calculated with MONASH 
and ATON codes. The ATON models of initial mass $2.5, 3~M_{\odot}$ and metallicity
$Z=8\times 10^{-3}$ are indicated with solid, black lines, whereas the ATON $1.5~M_{\odot}$,
$Z=4\times 10^{-3}$ evolutionary sequence is plotted with a solid, red track. The dotted,
magenta and green lines indicate the evolution of the MONASH models of metallicity
$Z=8\times 10^{-3}$ and initial mass, respectively, $3~M_{\odot}$ and $2.25~M_{\odot}$;
the MONASH model of mass $1.5~M_{\odot}$ and $Z=4\times 10^{-3}$ is indicated with a
dashed, blue line. Note that while the ATON tracks show the whole AGB evolution, 
including the thermal pulses, the MONASH lines are limited to the inter-pulse phases.}
\label{fmodel}
\end{figure*}

\subsection{Dust formation in the winds of AGB stars}
\label{dustmodel}
The growth of dust particles in the circumstellar envelope is described following
the pioneering explorations by the Heidelberg group \citep{gs85, gs99, 
fg01, fg02, fg06, zhukovska08}. The wind is assumed to expand isotropically
from the surface of the star; the velocity is held constant until gas molecules
enter the region where condensation into dust particles is favoured. As far as
a minimum amount of dust is formed, the results are independent of the adopted
initial velocity.

The model is based on the equation of momentum conservation and on a differential 
relation for the radial variation of the optical depth:

\begin{equation}
v{dv\over dr}=-{GM\over r^2}(1-\Gamma)
\label{eqv}
\end{equation}

\begin{equation}
{d\tau_L\over dr}=-\rho k{R^2\over r^2}
\label{eqtau}
\end{equation}

\noindent
$\Gamma={kL\over 4\pi cGM}$ accounts for the effects of radiation pressure on
dust grains, while $k$ is the extinction coefficient.

The description of the wind is completed by the equation giving the radial
variation of temperature and by the mass continuity relation:

\begin{equation}
T^4={1\over 2}T_{\rm eff}^4 \left[ 1-\sqrt{1-{R^2\over r^2}}+{3\over 2}\tau_L \right].
\label{eqt}
\end{equation}

\begin{equation}
\dot{M}=4\pi r^2 \rho v
\label{eqrho}
\end{equation}

\noindent
Here we focus on carbon-rich environments, where the surface carbon abundance exceeds oxygen.
In this case the two main dust species formed are solid carbon (C) and silicon carbide (SiC).
The condensation reactions by which the two species form are, respectively,
$C_2H_2 \rightarrow 2C(s)+H_2$ and $2Si+C_2H_2 \rightarrow 2SiC(s)+H_2$
\citep[see, e.g.][]{fg06}. The condensation
temperatures are 1100 K (C) and 1400 K (SiC). The growth rate of dust grains
($da_C/dt$ and $da_{SiC}/dt$) is proportional to the least abundant among the species 
involved in the condensation reactions, namely $C_2H_2$ and silicon. Considering the
high stability of CO and SiS molecules, we have

\begin{equation}
{da_C \over dt} \propto n_{C_2H_2} \propto (1-f_C)n_C-n_O-(1-f_{SiC})n_{Si}
\label{acar}
\end{equation}

\begin{equation}
{da_{SiC} \over dt} \propto (1-f_{SiC})n_{Si}-n_S
\label{asic}
\end{equation}

\noindent
In the above equations $f_C$ ($f_{SiC}$) indicates the fraction of
carbon (silicon) molecules condensed into dust particles.

SiC is the most stable species and forms at higher temperatures, 
thus closer to the surface of the star. Equation \ref{asic} indicates an upper limit
to the growth of SiC particles, reached when the fraction of silicon molecules
condensing into SiC approaches $f_{SiC} = 1-n_S/n_{SiC} \sim 0.55$. Because neither 
silicon nor sulphur undergo significant variations along the life of stars of intermediate
mass, this quantity remains constant during the AGB phase.

While the formation of SiC grains is important for the IR properties of the SED of AGB stars, 
it does not influence the dynamical status of the wind. This is because SiC particles are
extremely transparent to electromagnetic radiation and thus do not contribute to the
acceleration of the wind via radiation pressure.

Solid carbon is less stable than SiC, therefore the zone where the formation of carbon 
grains occurs is more external than the region where SiC particles begin to form. 
Because the surface carbon abundance is generally much higher than that of silicon, 
contrary to SiC, 
no saturation in the carbon dust formation process occurs.  The only exception is at the very 
beginning of the C-rich phase, when the carbon excess with respect to oxygen is very small, 
which prevents the formation of significant quantities of carbon dust (see equation 
\ref{acar}). 

However, because of the very large values attained by the extinction coefficient of solid 
carbon grains, the growth of these particles is commonly halted by the considerable 
expansion of the wind, which determines a fast drop in the gas density, according to 
equation \ref{eqrho}.

We may safely assume in this case (e.g., equation \ref{acar}) that the key 
quantity is the density of $C_2H_2$ molecules\footnote{This is consistent with mid-IR 
spectroscopic observations of Galactic and MCs AGB stars, which show that C$_{2}$H$_{2}$ 
is very abundant in their circumstellar envelopes (e.g., van Loon et al. 1999; Yang et al. 
2004; Lagadec et al. 2007; van Loon et al. 2008).}. The density of $C_2H_2$ in turn 
depends on the surface carbon mass fraction and on the density of the wind in the region 
where condensation occurs, i.e., the zone where the temperature is $T_C = 1100$ K; 
we refer to the density of this region as $\rho_C$.

Because the typical distance from the surface at which solid carbon forms is $\sim 10~R_*$, 
we may safely expand equation \ref{eqt} assuming that $R_*/r \ll 1$, to obtain 

\begin{equation}
(R_*/r)^2 \sim 4(T_C/T_{\rm eff})^4-4/3\tau_L
\label{approx}
\end{equation}

\noindent
The combination of equations \ref{eqrho} and \ref{approx} leads to:

\begin{equation}
\rho_C \sim {\dot M\over v} \left[ {T_C^4 \over L} - {\tau_L\over 12\pi \sigma R_*^2} \right]
\label{rhof}
\end{equation}

\noindent
The first term on the right hand side of equation \ref{rhof} is dominant. Therefore, the
number of $C_2H_2$ molecules available to form solid carbon is determined by the carbon 
mass fraction, X(C) and the $\dot M/L$ ratio.

We may therefore conclude that the winds of carbon stars have the following stratification:

\begin{itemize}

\item{SiC forms in an internal zone, at a temperature of $\sim 1400$ K. In most cases 
the growth of SiC particles is halted by saturation, which occurrs when the fraction of
gaseous silicon condensing into dust approaches $\sim 55 \%$ (see discussion in the 
previous section). The wind is modestly accelerated by formation of SiC, because this 
species is highly transparent to radiation.}

\item{Solid carbon particles form in more external regions, where the temperature is
$\sim 1100$K. In this case the wind is greatly accelerated owing to the high values of
the extinction coefficient of carbon dust. The factors mostly relevant to formation of
solid carbon are the $\dot M/L$ ratio and the surface carbon mass fraction.}

\end{itemize}

\subsection{Synthetic spectra}
\label{spectramodel}
To determine the magnitudes in the {\it Spitzer} bands used in the present analysis 
we follow the same
approach used in \citet{ventura15a} and discussed in detail in \citet{flavia15a}.
Based on the values of mass, luminosity, effective temperature, mass-loss rate and 
surface chemical composition calculated from the MONASH and ATON codes, we apply
the model of dust formation described in Section \ref{dustmodel} to determine the size of the dust 
grains formed and the optical depth (here we use the value at $10~\mu$m, $\tau_{10}$).

These ingredients are used by the code DUSTY \citep{dusty} to calculate the synthetic 
spectra of each selected point along the evolutionary sequence. The magnitudes in the 
various bands are obtained by convolution with the appropriate transmission curves.
DUSTY needs as input parameters the effective temperature of the star, the
radial profile of the density of the gas and the dust composition of the wind, in terms
of the percentage of the various species present and of the size of the dust particles
formed. All these quantities are known based on the results of stellar evolution and of
the description of the wind.

\section{Carbon stars in the Magellanic Clouds: evolution and dust properties}
\label{evolution}
We focus now on the stars that account for most of the C-star population in the LMC and SMC. 
For the LMC we consider models of initial mass $\sim 2.5-3~M_{\odot}$ and metallicity 
$Z=8\times 10^{-3}$. To compare models with the same core mass, while in the ATON case
we show models of initial mass $2.5~M_{\odot}$ and $3~M_{\odot}$, in the MONASH case we
present the  $2.25~M_{\odot}$ and $3~M_{\odot}$ models; this is due to a slight difference
in the mass vs core mass relationships found with the two codes in the range of masses
close to the limit to experience the helium flash.
For the SMC we analyse the evolution of $\sim 1.5~M_{\odot}$ stars of metallicity 
$Z=4\times 10^{-3}$. 
 
\subsection{Physical properties of carbon stars} 
The evolution of these models is shown in Fig. \ref{fmodel}, where we plot the luminosity,
effective temperature, mass-loss rate and surface carbon abundance as a function of time
from the beginning of the thermally-pulsing AGB phase. 
 
The top, left panel of Fig. \ref{fmodel} shows that the ATON and MONASH models evolve 
at similar luminosities, an indication that the core masses are similar. 
Note that in this case we do not expect any effect of convection modelling on the
behaviour of luminosity, because stars in this range of mass do not experience 
HBB. This is at odds with the analysis by \citet{ventura15a}, where the
ATON and MONASH models of $5-6~M_{\odot}$ stars were shown to evolve at different luminosities.
The only meaningful difference difference is the duration of the whole AGB phase, which 
is systematically longer in the MONASH models; we will come back to this point shortly. 
The duration of the AGB phase is not monotonic with mass: the $2.5~M_{\odot}$ model
evolves for a longer time compared to their $1.5~M_{\odot}$ and $3~M_{\odot}$ counterparts. 
This is because the relationship between the core mass at the beginning of the AGB phase 
and the initial mass of the stars has a minimum at $\sim 2.5~M_{\odot}$, which makes the 
growth of the luminosity of the $2.5~M_{\odot}$ model in the initial AGB phases slower 
than in the other cases.

In the phases before the C-star phase is reached, the effective temperatures of models 
of the same mass are similar in the MONASH and ATON cases (see top, right panel of
Fig. \ref{fmodel}). After the C/O ratio exceeds unity, the effects of the different 
low-temperature opacities takes over: the ATON models, owing to the expansion of the 
external regions triggered by the increase in the opacity, evolve to lower temperatures, 
down to $\sim 2000$ K, whereas in the MONASH models we find $T_{\rm eff} > 2400$ K. This effect
is extensively discussed in \citet{vm09, vm10}.

The difference in the surface temperature has a direct effect on the rate of mass loss.
As shown in the bottom, left panel of Fig. \ref{fmodel}, $\dot{M}$ exceeds
$10^{-4}~M_{\odot}$/yr in the ATON models of $Z=8\times 10^{-3}$, whereas in the MONASH  
case it is $\sim 3-4$ times smaller; in the $1.5~M_{\odot}$ model of metallicity 
$Z=4\times 10^{-3}$ the difference is smaller
($2\times 10^{-5}~M_{\odot}$/yr and $3\times 10^{-5}~M_{\odot}$/yr in the MONASH and
ATON models, respectively). This is not only due to the more expanded, hence less 
gravitationally bound, configurations of lower $T_{\rm eff}$'s models; a further reason
for the higher $\dot M$'s of the ATON models is the use of the formulation by
\citet{wachter02, wachter08}, which favours large rates of mass loss in carbon rich,
low $T_{\rm eff}$ stars. The rate of mass loss in the ATON models is
generally larger than in the MONASH cases also during the early AGB phases. 
This is due to the differences between the \citet{blocker95} and \citet{VW93}
formulae for M stars. This is not a major problem here because little mass is lost
during the oxygen-rich phase of these stars and, as will be shown in the following,
the amount of dust formed during these evolutionary phases is negligible in comparison
to the late evolutionary phases, when the stars are carbon-rich.

\subsection{The change in the surface chemistry}
Concerning the surface chemistry, we show in the bottom, right panel of 
Fig. \ref{fmodel} the variation of the carbon abundance in the same models shown in 
the other panels. The most relevant difference between the results found with the two
evolution codes is found in the $3~M_{\odot}$ case, for which the 
MONASH and ATON models reach final carbon mass fractions of $X_C=0.0185$ and
$X_C=0.01$, respectively. This situation is reversed in lower mass models; indeed in
the $1.5~M_{\odot}$ ATON case the final surface carbon is $X_C \sim 0.007$, a factor of 2
higher than MONASH. This is because the MONASH models do not show very efficient
dredge-up at the minimum mass for C-star production, a problem that has been discussed
extensively in \citet{kamath12} and \citet{karakas10b}. The 1.5$\Msun$ model only dredges 
up $0.0096~M_{\odot}$ of material from the He-intershell, with a minimum core mass for TDU of 
0.6$~\Msun$. \citet{kamath12} found  that the minimum mass for TDU in LMC AGB stars should 
be lower, at 0.58$~\Msun$, indicating that convective overshoot should probably be applied 
\citep[see also, e.g.,][]{marigo99}. To reach surface carbon mass fractions 
$X_C > 0.005$ an extra-mixing zone extending over two pressure scale heights below
the base of the convective envelope needs to be applied. This is entirely consistent with
the analysis by \citet{kamath12}, who invoked an overshoot of similar extension to
match the observations of AGB stars in MCs clusters.

These dissimilarities between the results obtained with the two evolution codes are 
not surprising, given the considerable sensitivity of the carbon enrichment of the 
surface layers to the details of the treatment of convective borders, as thoroughly 
documented in the literature \citep{karakas14}.

\subsection{Dust formation in the circumstellar envelope}
\label{dustgrain}

\begin{figure*}
\begin{minipage}{0.47\textwidth}
\resizebox{1.\hsize}{!}{\includegraphics{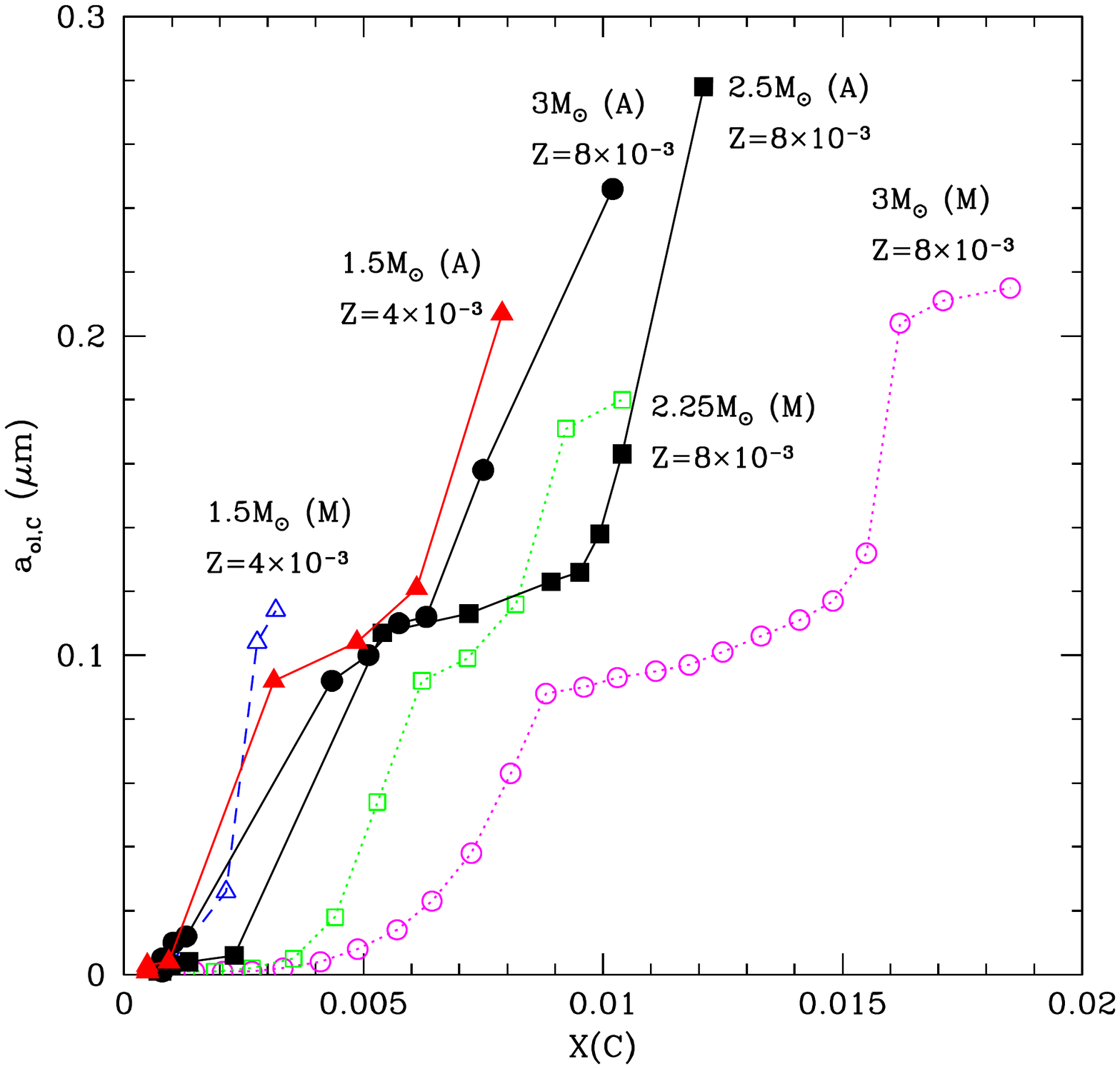}}
\end{minipage}
\begin{minipage}{0.47\textwidth}
\resizebox{1.\hsize}{!}{\includegraphics{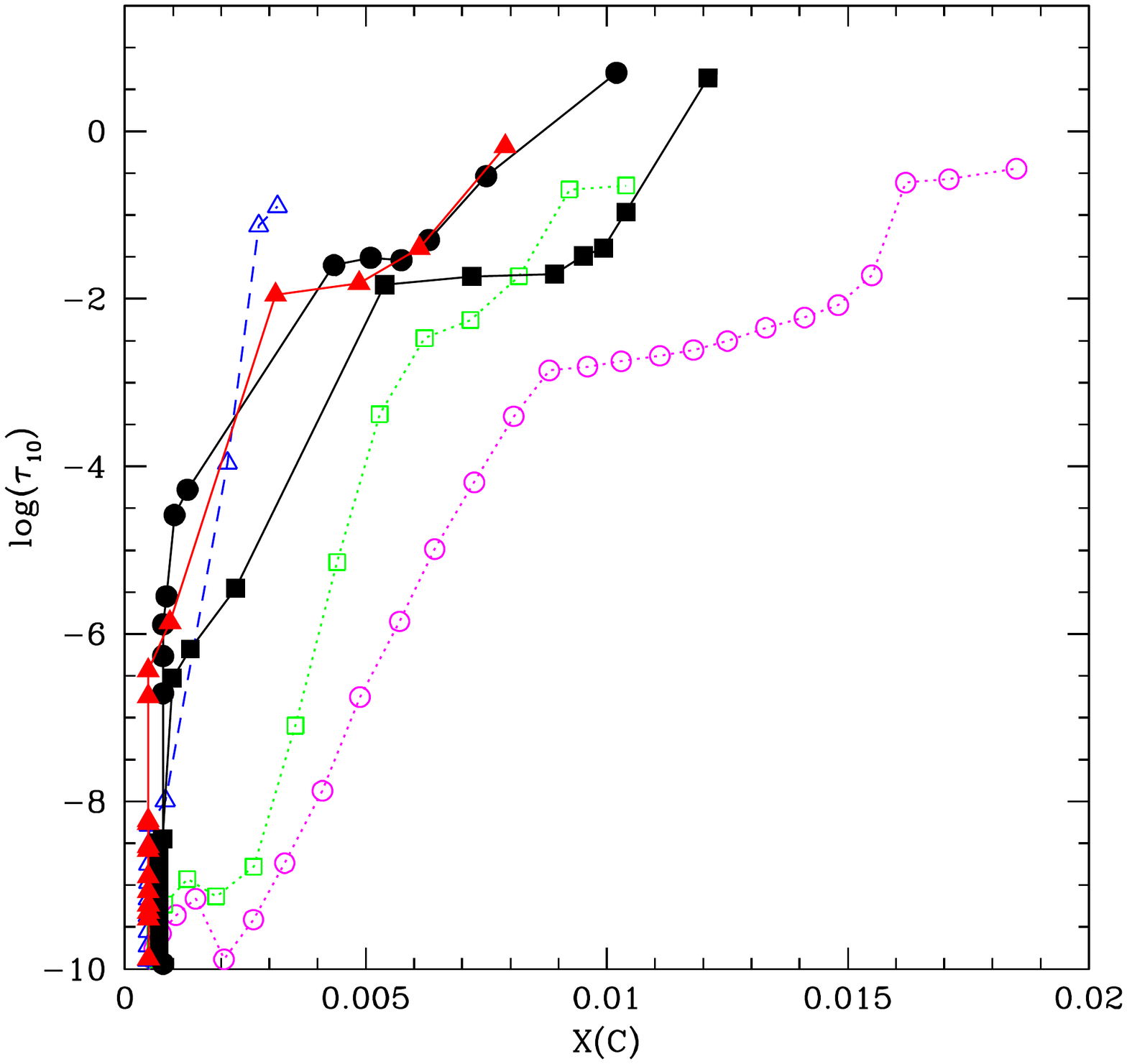}}
\end{minipage}
\vskip-50pt
\caption{The variation of the size of the olivine (oxygen-rich phases) and solid
carbon (carbon stars) dust grains (left panel) and of the optical depth at 
$10 \mu$m (right) in the same models shown in Fig.~\ref{fmodel}. On the abscissa 
we show the surface carbon abundance, increasing during the AGB phase.
Symbols along the tracks in the left panel indicate the initial mass and whether
the MONASH (M) or ATON (A) code was used for the calculation.
}
\label{ftau}
\end{figure*}

The changes in the spectral energy distribution of AGB stars are correlated with 
evolution properties. This reflects the variation of the main physical parameters 
(e.g., mass loss) and of the quantity of dust in the circumstellar envelope.

Fig.~\ref{ftau} shows the evolution of the size of the dust grains formed (left panel)
and the optical depth at wavelength $\lambda=10~\mu$m, $\tau_{10}$, for the same
models shown in Fig.~\ref{fmodel}. For clarity, we only show the dimension of
the dust particles formed in greater quantities, which provide the dominant contribution
to the acceleration of the wind, i.e., olivine (oxygen-rich phases) and
solid carbon (C-stars). $\tau_{10}$ is plotted on a logarithmic scale, which allows 
us to discuss the early AGB phase, which is otherwise undetectable on a linear scale 
plot. On the abscissa we show the surface mass fraction of carbon, which increases 
during the AGB phase.

The following results are in common between the MONASH and ATON models:

(i) little dust formation occurs until the C/O ratio is close to unity. 
As shown in Fig. \ref{ftau}, the olivine grains barely exceed a nanometer 
size and the optical depth is very small. The SED of the star is unaffected by dust 
during these phases. This holds independently of mass and metallicity.

\begin{figure*}
\begin{minipage}{0.32\textwidth}
\resizebox{1.\hsize}{!}{\includegraphics{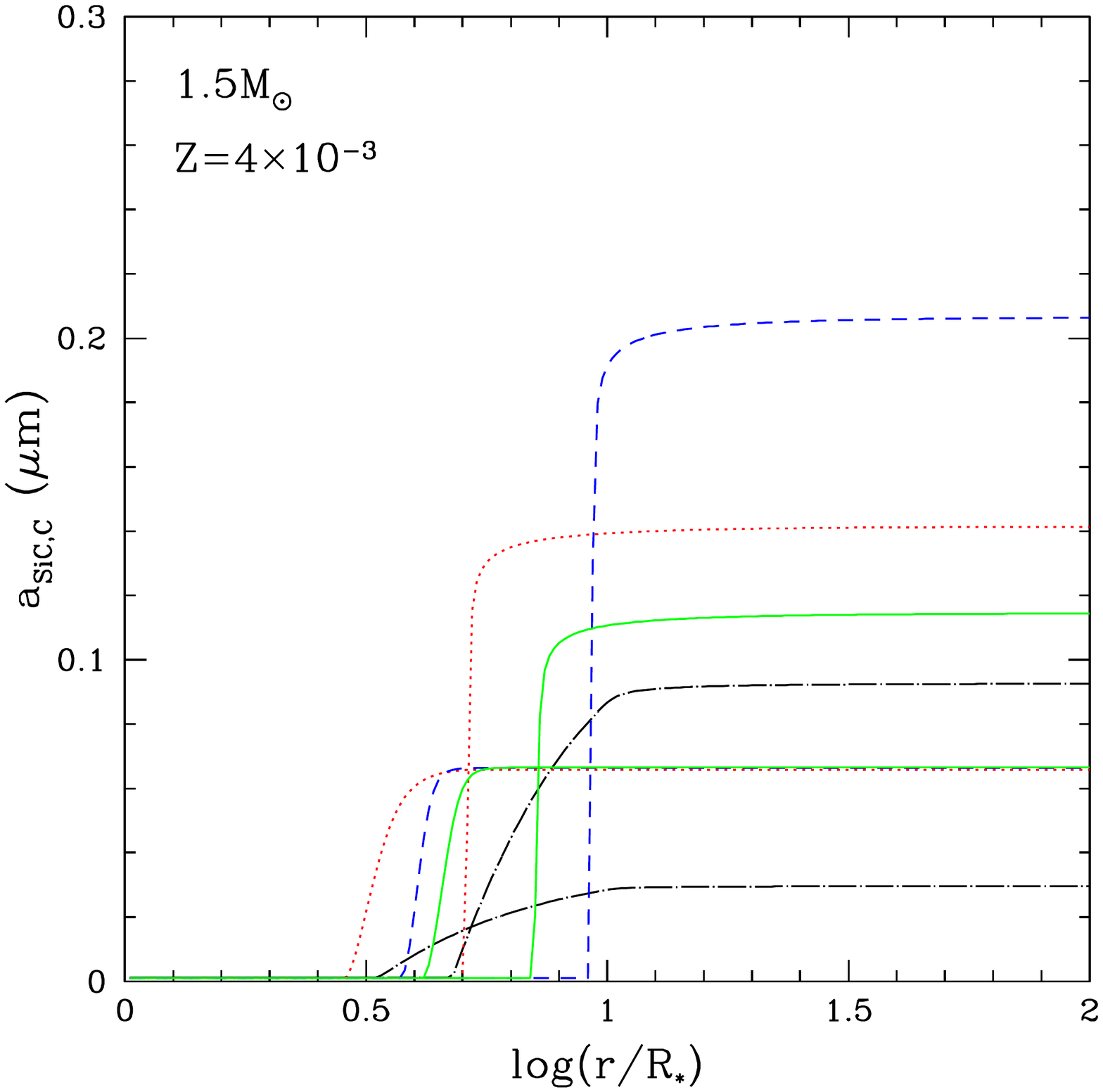}}
\end{minipage}
\begin{minipage}{0.32\textwidth}
\resizebox{1.\hsize}{!}{\includegraphics{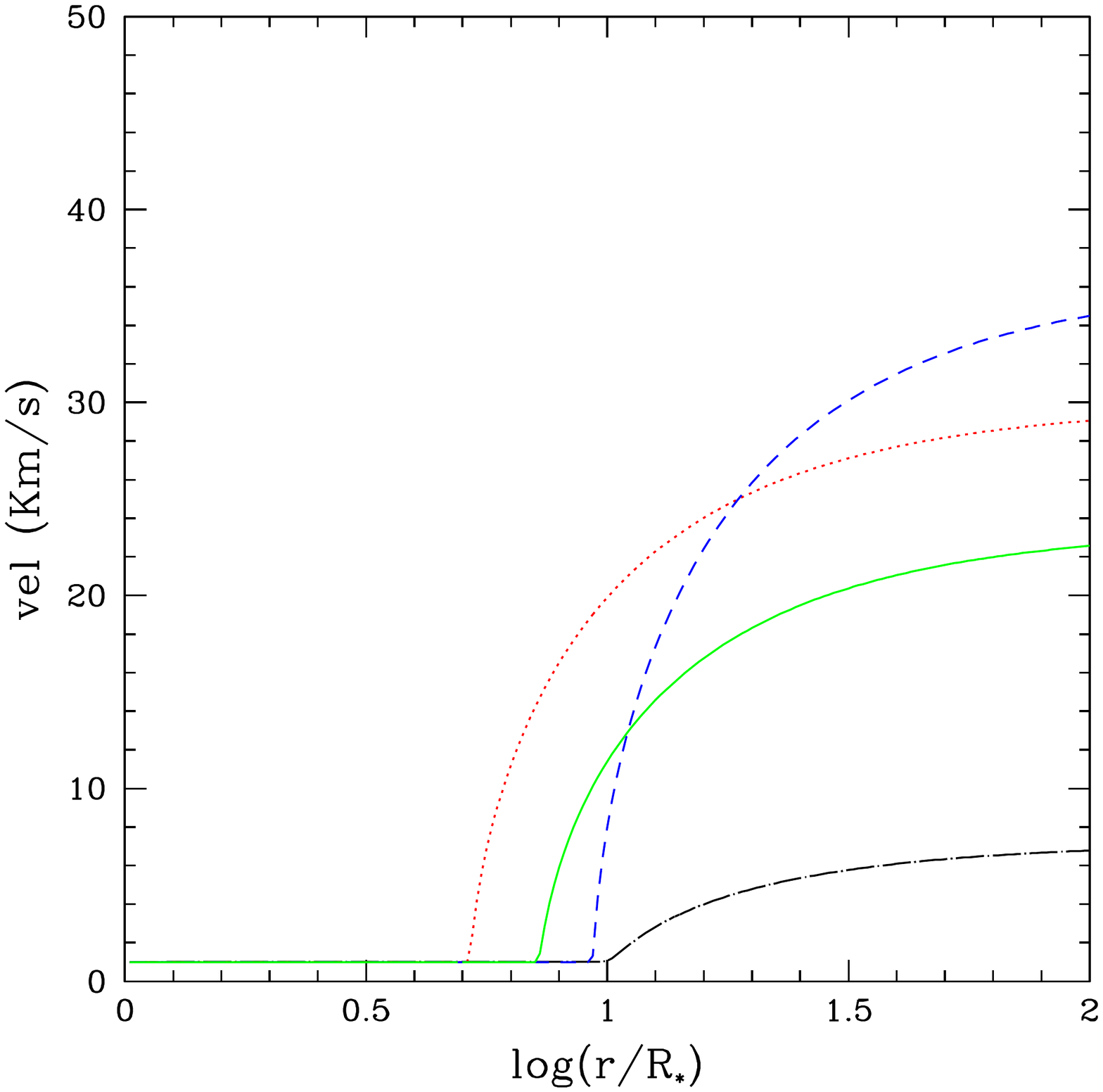}}
\end{minipage}
\begin{minipage}{0.32\textwidth}
\resizebox{1.\hsize}{!}{\includegraphics{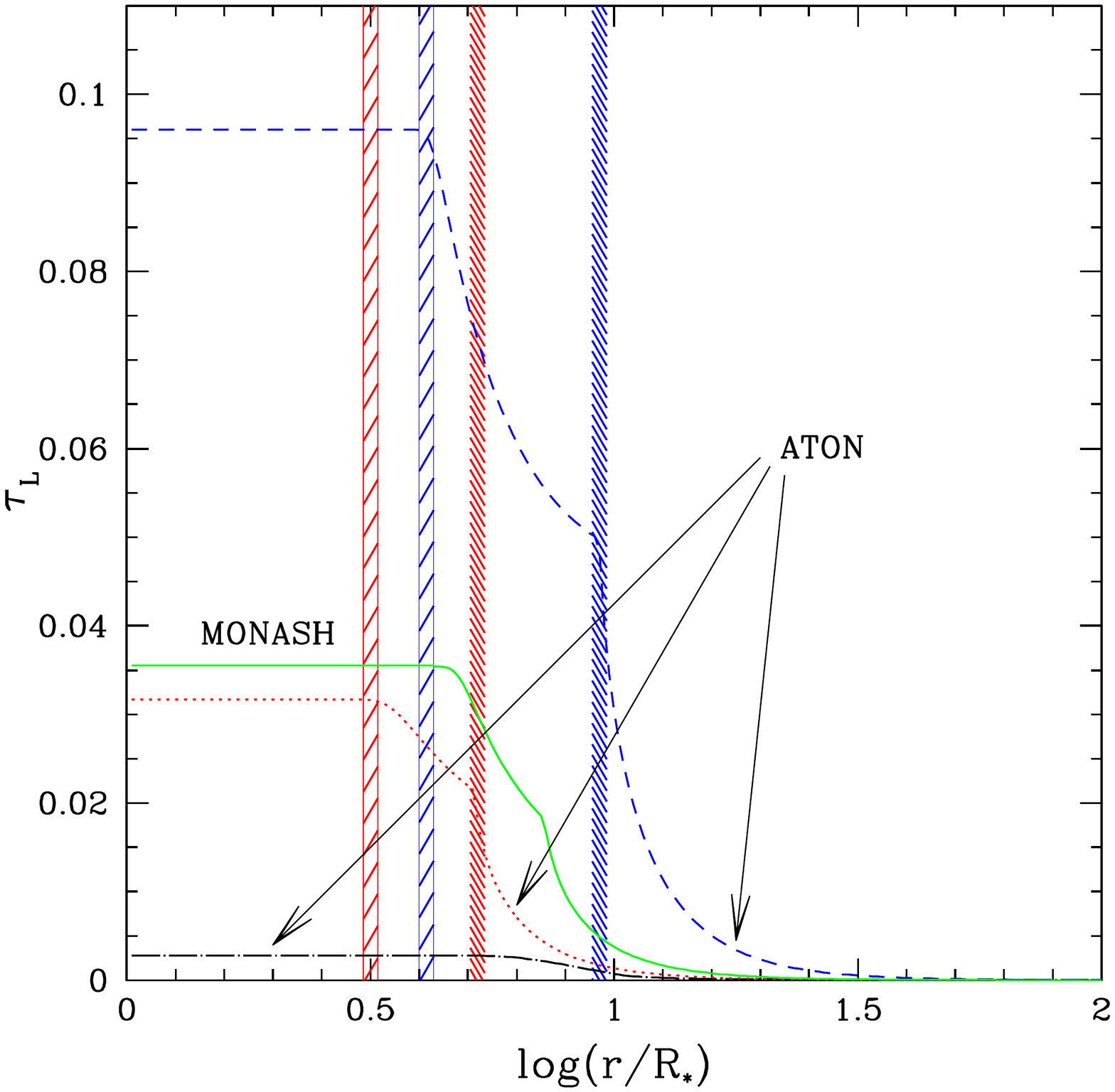}}
\end{minipage}
\vskip-40pt
\begin{minipage}{0.32\textwidth}
\resizebox{1.\hsize}{!}{\includegraphics{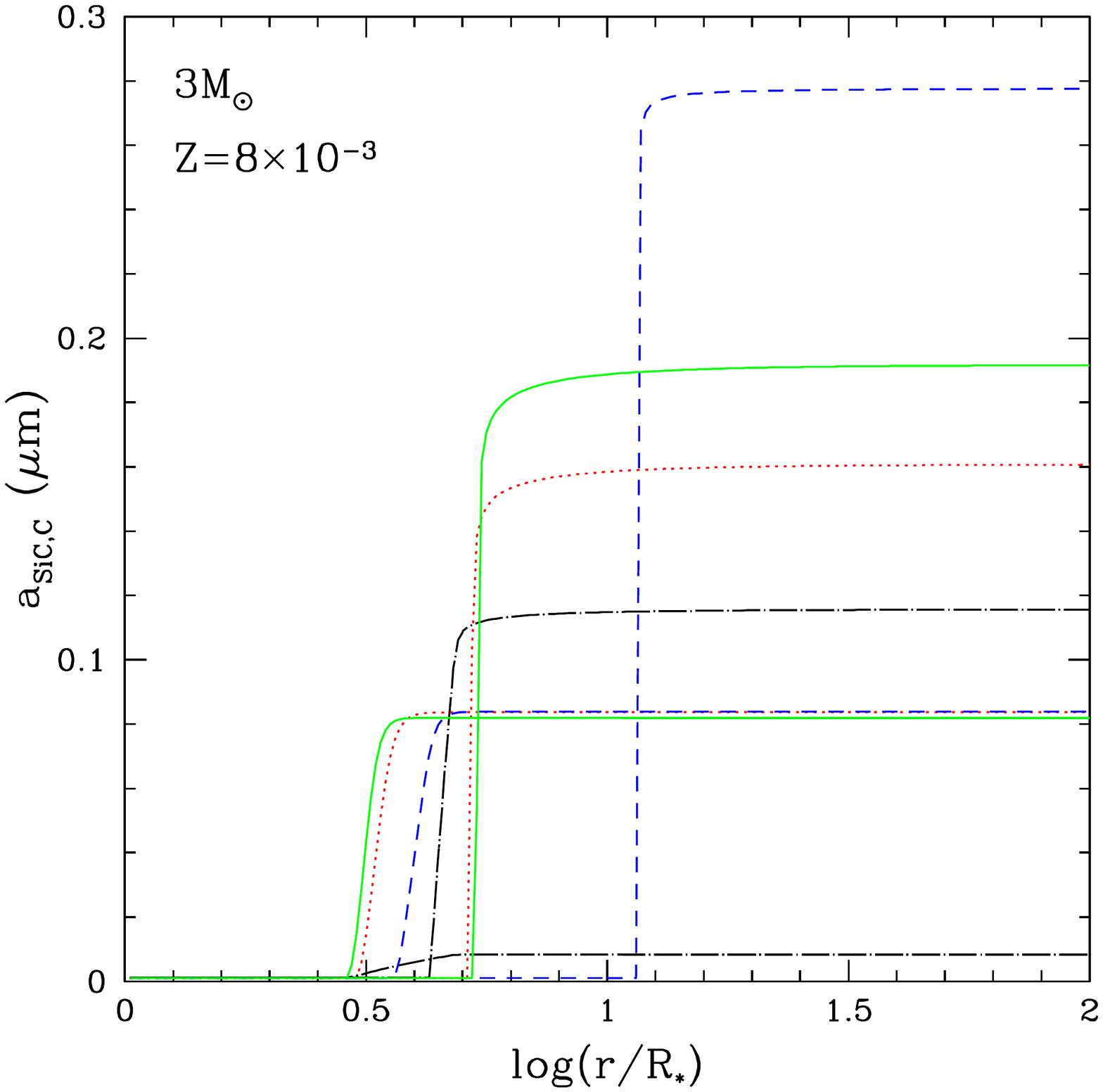}}
\end{minipage}
\begin{minipage}{0.32\textwidth}
\resizebox{1.\hsize}{!}{\includegraphics{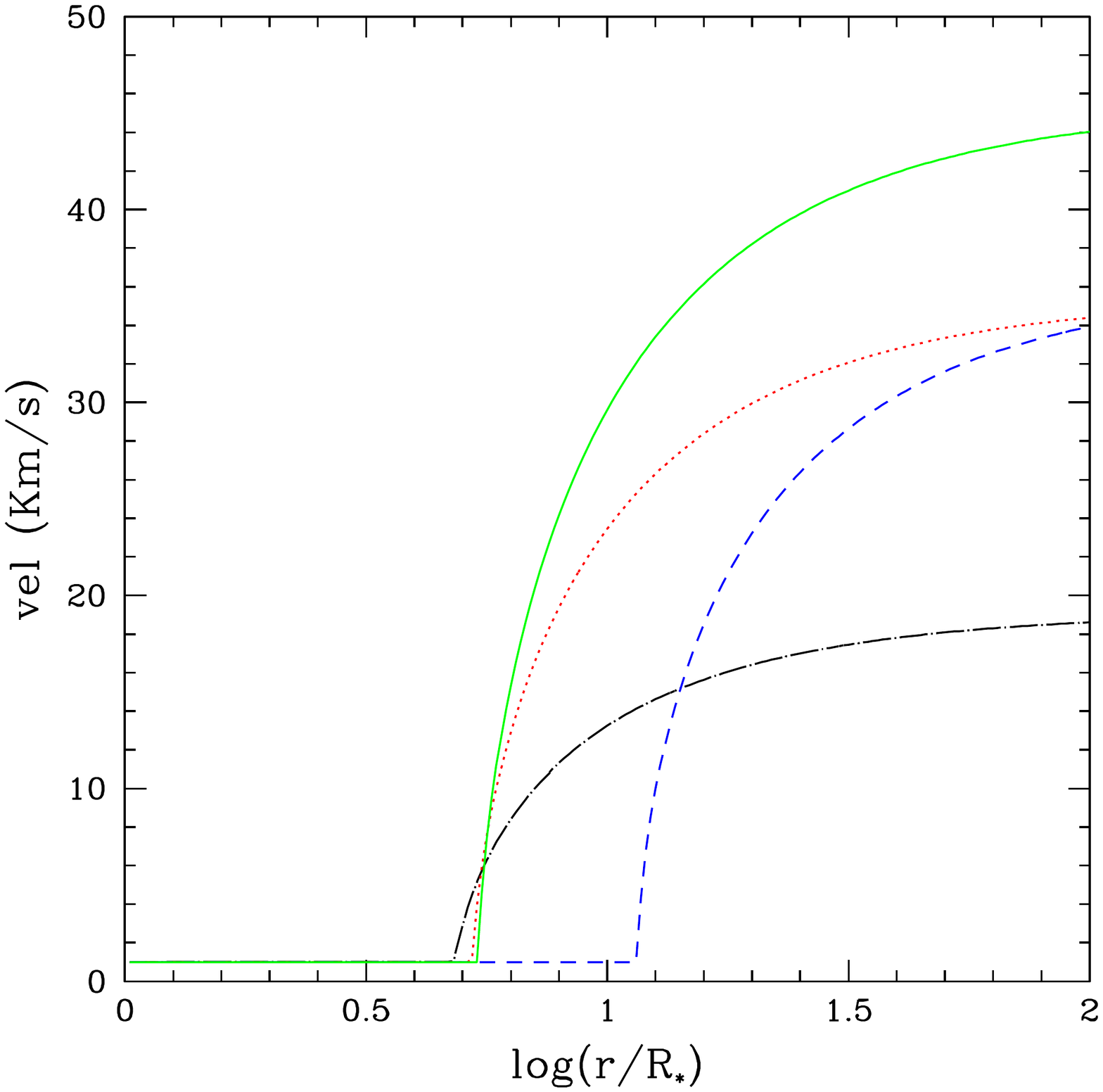}}
\end{minipage}
\begin{minipage}{0.32\textwidth}
\resizebox{1.\hsize}{!}{\includegraphics{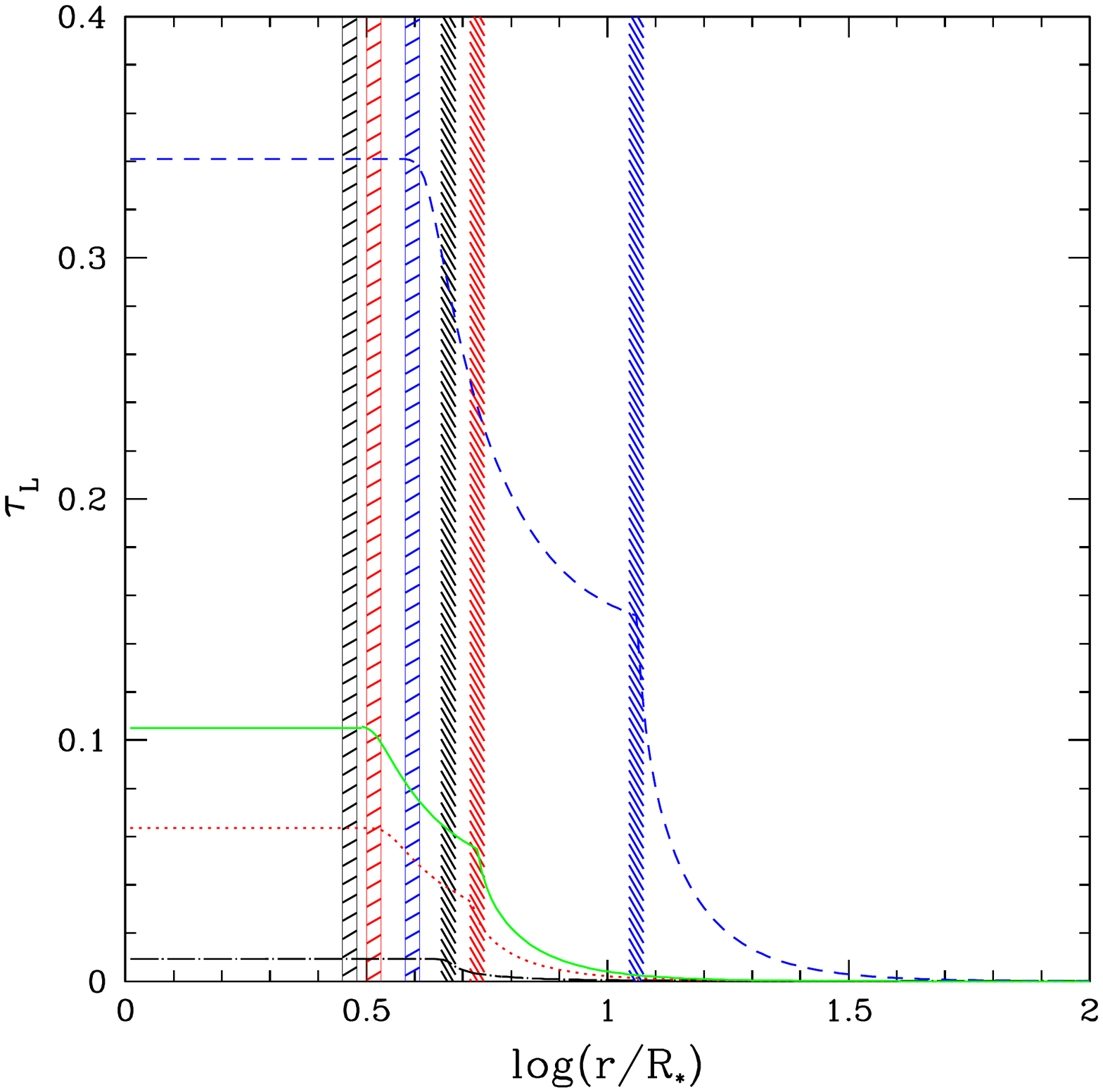}}
\end{minipage}
\vskip-30pt
\caption{The radial stratification of the size of SiC and carbon grains (left panels),
wind velocity (middle) and $\tau_L$ (right) in different phases during the AGB evolution
of models of initial mass $1.5~M_{\odot}$ (top) and $2.5~M_{\odot}$ (bottom). On the
abscissa we show the distance from the surface of the star, on a logarithmic scale, 
measured in stellar radii units. Black, dotted--dashed, red dotted  and dashed, blue 
tracks refer to ATON models taken, respectively, at the beginning of the C-star phase, 
in an intermediate stage after becoming C-star and at the end of the AGB evolution.
Solid, green lines correspond to MONASH models of initial mass $1.5~M_{\odot}$ (top) and 
$3~M_{\odot}$ (bottom), taken in the final evolutionary phases. The shaded regions outlined
in the right panels indicate the zones where formation of SiC and solid carbon dust 
occurs.
}
\label{fstrat}
\end{figure*}

(ii) The quantity of carbon dust formed increases during the subsequent C-star evolution,
as confirmed by the rise of the grain size and $\tau_{10}$, shown in Fig.~\ref{ftau}.
This can be understood based on the discussion in Section \ref{dustmodel}, because
both factors affecting formation of carbonaceous dust, namely the surface 
carbon abundance and the $\dot M/L$ ratio, increase after becoming C-rich (see 
Fig.~\ref{fmodel}). 

(iii) The amount of dust formed in the winds of 
the $\sim 1.5~M_{\odot}$ star is significantly smaller than in their counterparts of higher 
mass. This is because the $2.5-3~M_{\odot}$ models, particularly during
the final AGB phases, attain a higher surface carbon abundance, which in turn leads to 
lower effective temperatures and higher rates of mass loss;
this can be clearly seen in Fig.~\ref{fmodel}.

\subsection{Thermodynamic and chemical structure of the wind}
To understand how the conditions in the wind evolve, we show in Fig.~\ref{fstrat} 
the stratification of the circumstellar envelope in three distinct stages of the
carbon star phase of the $1.5~M_{\odot}$, $Z=4\times 10^{-3}$ (top panel) and the 
$2.5~M_{\odot}$, $Z=8\times 10^{-3}$ ATON models.
The plots illustrate conditions at the beginning of the AGB phase,  the very end of the
AGB evolution and an intermediate phase. In the same figure we also show the MONASH 
models of mass $1.5~M_{\odot}$ and $3~M_{\odot}$ during the final AGB phases. The 
panels of the figure show the radial distribution of the grains size of SiC and solid 
carbon, the velocity of the wind, and $\tau_L$, the quantity entering equations \ref{eqtau} 
and \ref{eqt}.

Formation of SiC occurs in an internal zone,
$\sim 2-3$ stellar radii away from the surface, where the temperature is $T \sim 1400$ K. 
Because SiC is highly transparent to stellar radiation, the velocity of the wind stays 
constant in the region of SiC formation. At the very beginning of the carbon star phase 
SiC is produced in modest quantities, with grains of a few nanometers in size. In more 
advanced phases saturation condition occurs, the dimension of SiC grains grow to 
$0.065 \mu$m and $0.081 \mu$m in the $1.5~M_{\odot}$ and $2.5~M_{\odot}$ cases, 
respectively. This is because silicon in the envelope is either locked into SiS molecules 
or condensed into SiC particles. The difference in the SiC grain size is due to the 
dependence of the surface silicon abundance on initial metallicity.

\begin{figure*}
\begin{minipage}{0.47\textwidth}
\resizebox{1.\hsize}{!}{\includegraphics{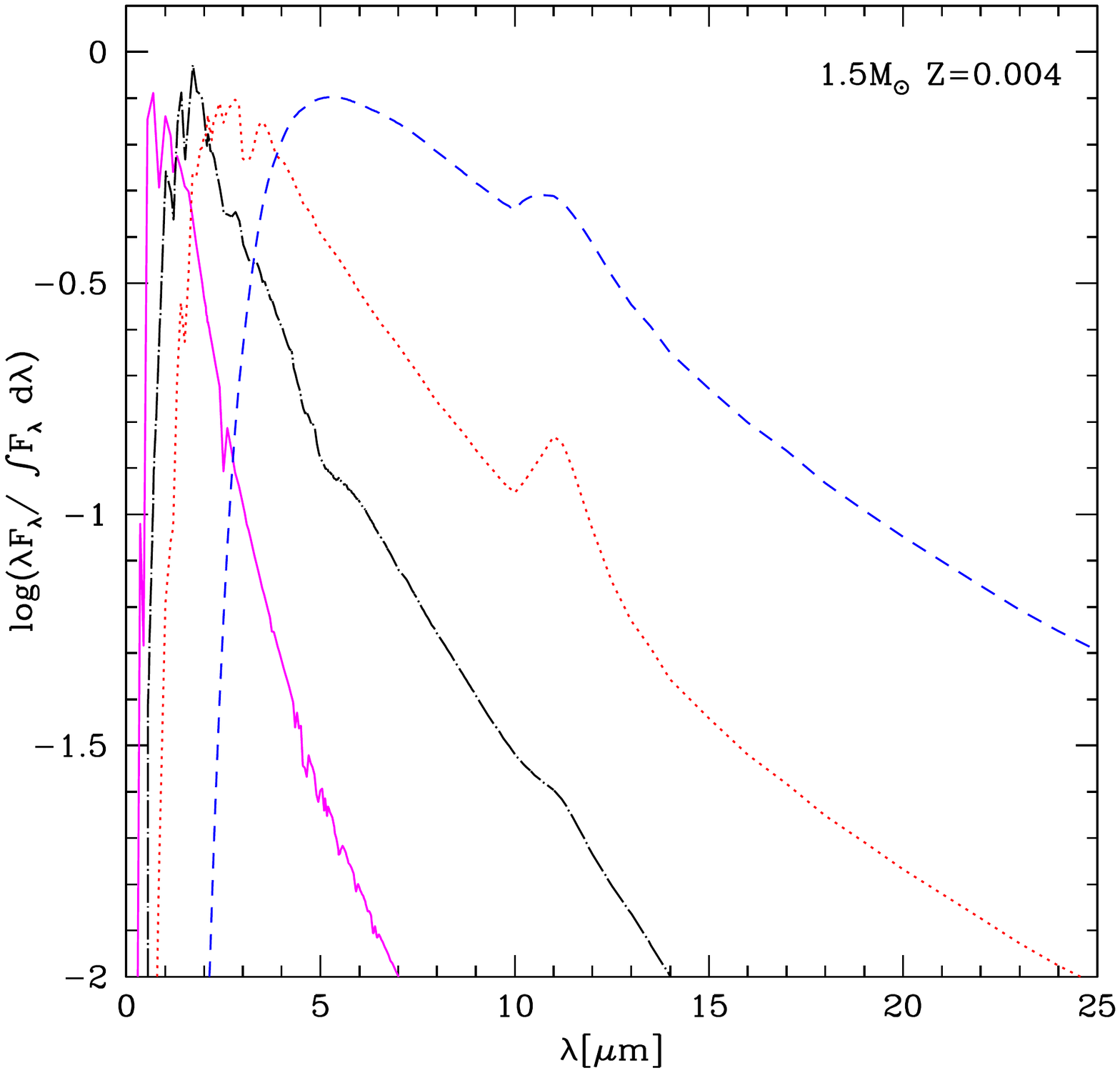}}
\end{minipage}
\begin{minipage}{0.47\textwidth}
\resizebox{1.\hsize}{!}{\includegraphics{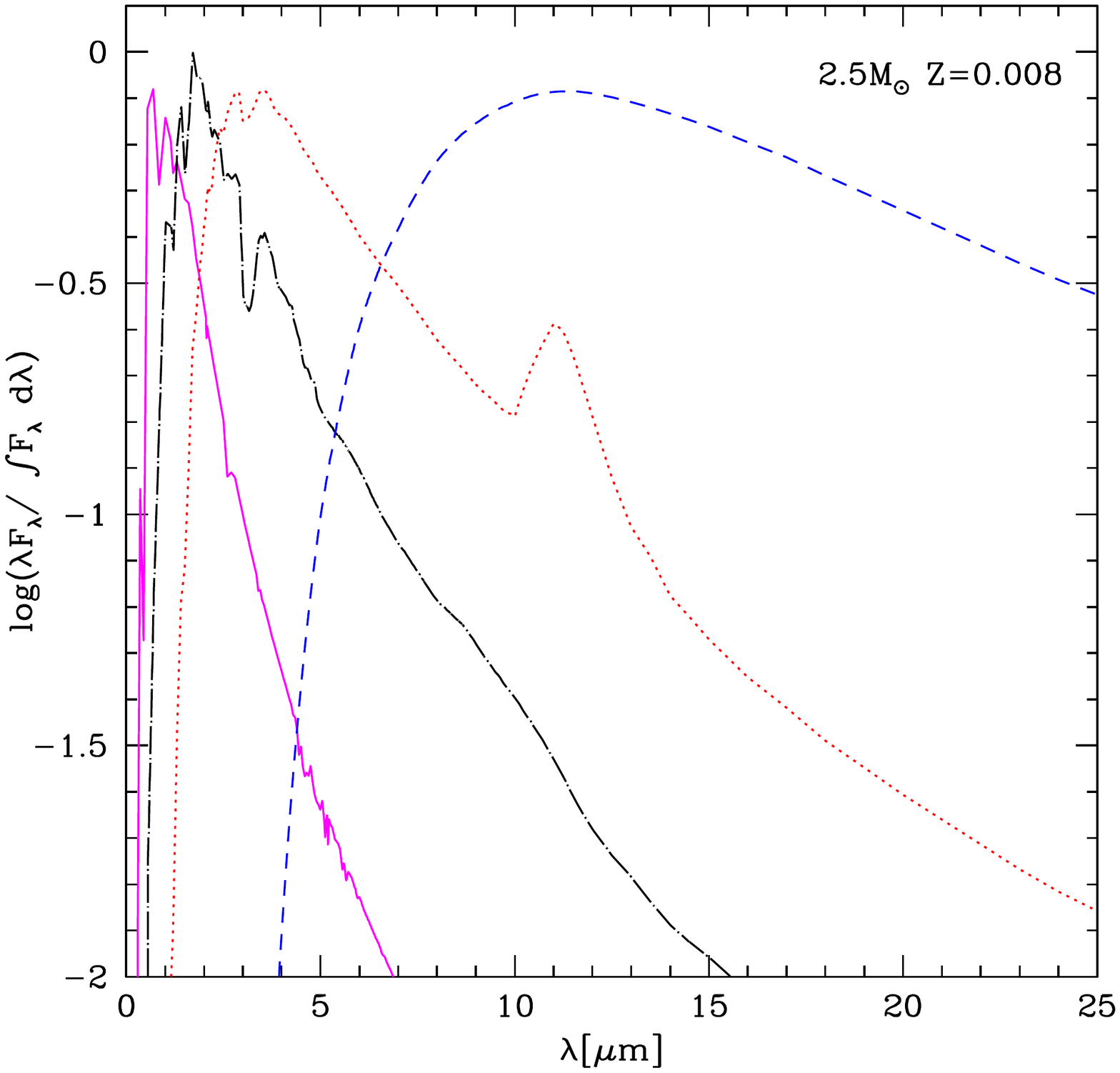}}
\end{minipage}
\vskip-40pt
\caption{The evolution of the SED of low-mass stars
evolving to the C-star stage during the AGB phase. The solid, magenta lines correspond
to the SED at the beginning of the AGB phase, when the stars are still oxygen-rich.
Dotted--dashed, dotted and dashed tracks correspond to the same model as in 
Fig.~\ref{fstrat}. 
}
\label{fsed}
\end{figure*}

Formation of carbon grains occurs $\sim 7-10$ stellar radii away from the surface, 
where the temperature is $T\sim 1100$ K. Carbon is produced in greater quantities compared 
to SiC, owing to the higher availability of carbon molecules in comparison to silicon. 
The formation of carbon dust is accompanied by the acceleration of the wind. The location 
of the carbon condensation zone is gradually shifted outwards during the AGB evolution; 
this is because the slope of the T(r) relation is smaller for lower effective temperatures
and the higher $\tau_L$ also result in pushing out the region where 
$T\sim 1100$ K (see equation \ref{eqt}). The typical size of the carbon grains formed, 
$a_C$, is of the order of $0.1-0.2~\mu$m and increases during the AGB evolution. In the 
final evolutionary phases of the $2.5~M_{\odot}$ model we find $a_C \sim 0.3~\mu$m, 
whereas we get a maximum dimension $a_C \sim 0.2~\mu$m in the $1.5~M_{\odot}$ case.

In the comparison between the SiC and solid carbon dust formed, we note that
the amount of SiC, after saturation conditions occur, stays constant during the
AGB evolution. On the other hand larger and larger quantities of solid carbon grains
are formed as the star accumulates carbon in the surface regions. Therefore, we expect
that the SiC/C ratio decreases during the AGB evolution (see Fig.~\ref{fsed} and section 6). 

The velocity reached by the wind also increases during the evolution of these stars,
spanning the range $10-50$ km/s.

\subsection{An overview of MONASH and ATON results}
The comparison between the MONASH and ATON results allows us to study
the main factors affecting carbonaceous dust formation in AGB stars.

The evolution of the infrared properties of these stars is strongly related to the 
amount of dust formed. 
As discussed in section \ref{dustmodel}, formation of solid carbon is determined
by the surface carbon abundance and the $\dot M/L$ ratio. The former is relevant 
at the beginning of the C-star phase, as confirmed by the steep slope of the 
$a_C$ vs X(C) and $\tau_{10}$ vs X(C) relations shown in Fig.~\ref{ftau}. This is
because the growth rate of carbon dust is proportional to the density of 
$C_2H_2$ molecules, which in turn, is given by the carbon excess with respect to
oxygen (see equation \ref{acar}). For small carbon abundances even a small change
of X(C) favours a high percentage increase in $n_{C_2H_2}$, which results in a
significant rise in the growth rate of carbon grains. After the surface carbon exceeds
$X(C) \sim 5\times 10^{-3}$ the relative weight of this factor diminishes, because any 
additional change will result in a smaller percentage increase in the number density
of $C_2H_2$ particles; the $\dot M /L$ ratio plays a dominant role here.

Fig~\ref{ftau} shows that at the end of the AGB evolution higher values of $\tau_{10}$ are
reached in the ATON models. For $M=1.5~M_{\odot}$ we find $\tau_{10} = 0.1$ in the
MONASH model, whereas we obtain $\tau_{10} = 0.7$ in the ATON case. 
In this case the star reaches relatively small carbon abundances, $X(C) < 7\times 10^{-3}$.
Following the above discussion, we know that this is the domain most affected by the 
carbon content. The ATON model produces more dust than MONASH (see left panel of 
Fig.~\ref{ftau}), owing to the higher availability of carbon in the surface regions, 
as shown in the bottom--right panel of Fig.~\ref{fmodel}. The infrared 
emission is consequently more intense in the ATON case.

Turning to models of higher mass and metallicity, we find that the $3~M_{\odot}$ MONASH 
model evolves to $\tau_{10} \sim 1$, while the $2.5~M_{\odot}$ ATON model reaches 
$\tau_{10} \sim 5$ at the end of the AGB phase. 
Stars of this mass undergo more TDU events, thus becoming more C-rich
during the AGB evolution. The rate of mass loss plays a dominant role here in
affecting dust formation. This is confirmed by the comparison between the
MONASH model of $3~M_{\odot}$ with the ATON, $2.5~M_{\odot}$ case: despite the former
experiences a larger carbon enrichment (see Fig.~\ref{fmodel}), the latter reaches higher 
optical depths in the final phases, owing to the higher rate of mass loss experienced.

\begin{figure*}
\begin{minipage}{0.47\textwidth}
\resizebox{1.\hsize}{!}{\includegraphics{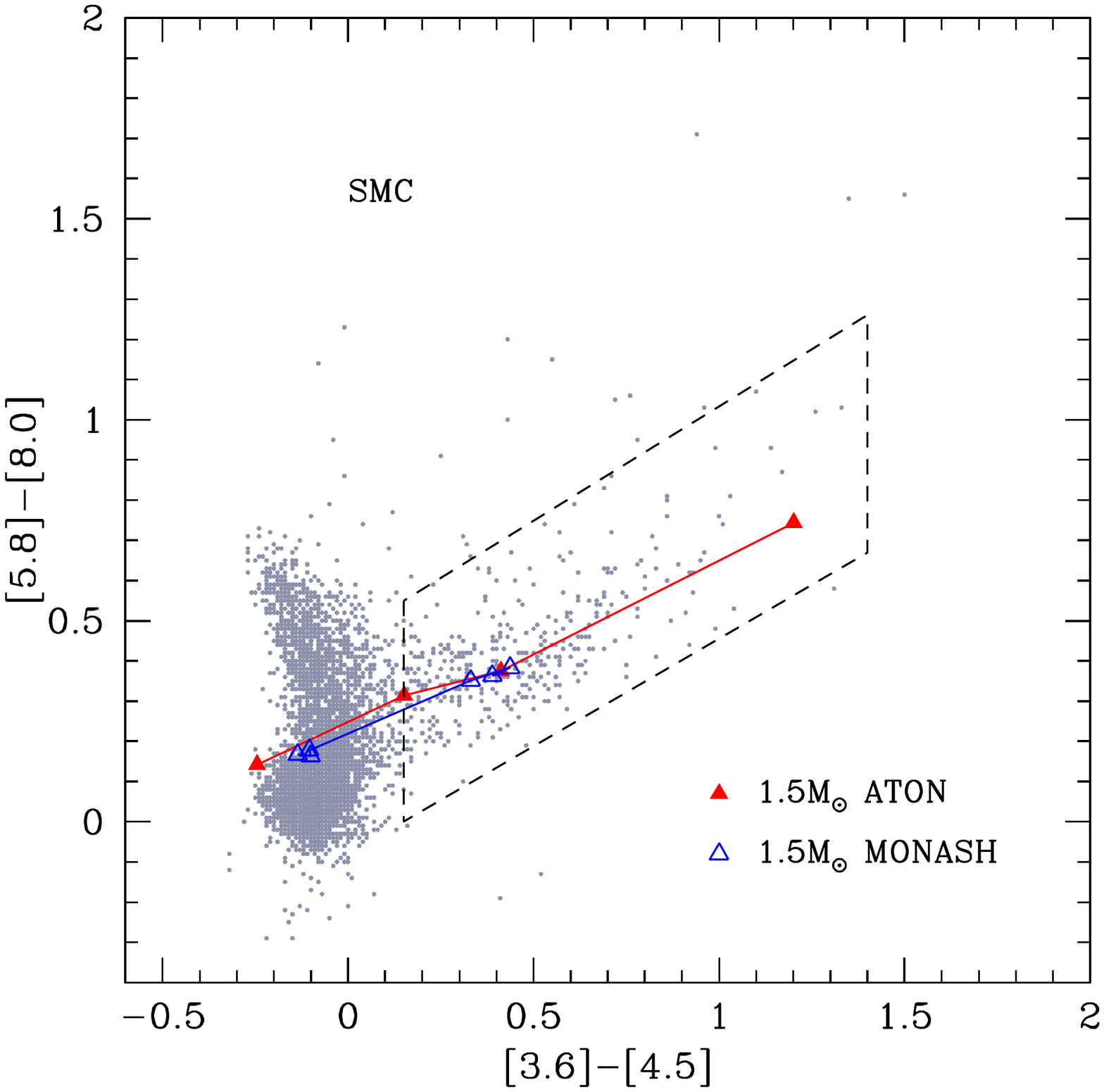}}
\end{minipage}
\begin{minipage}{0.47\textwidth}
\resizebox{1.\hsize}{!}{\includegraphics{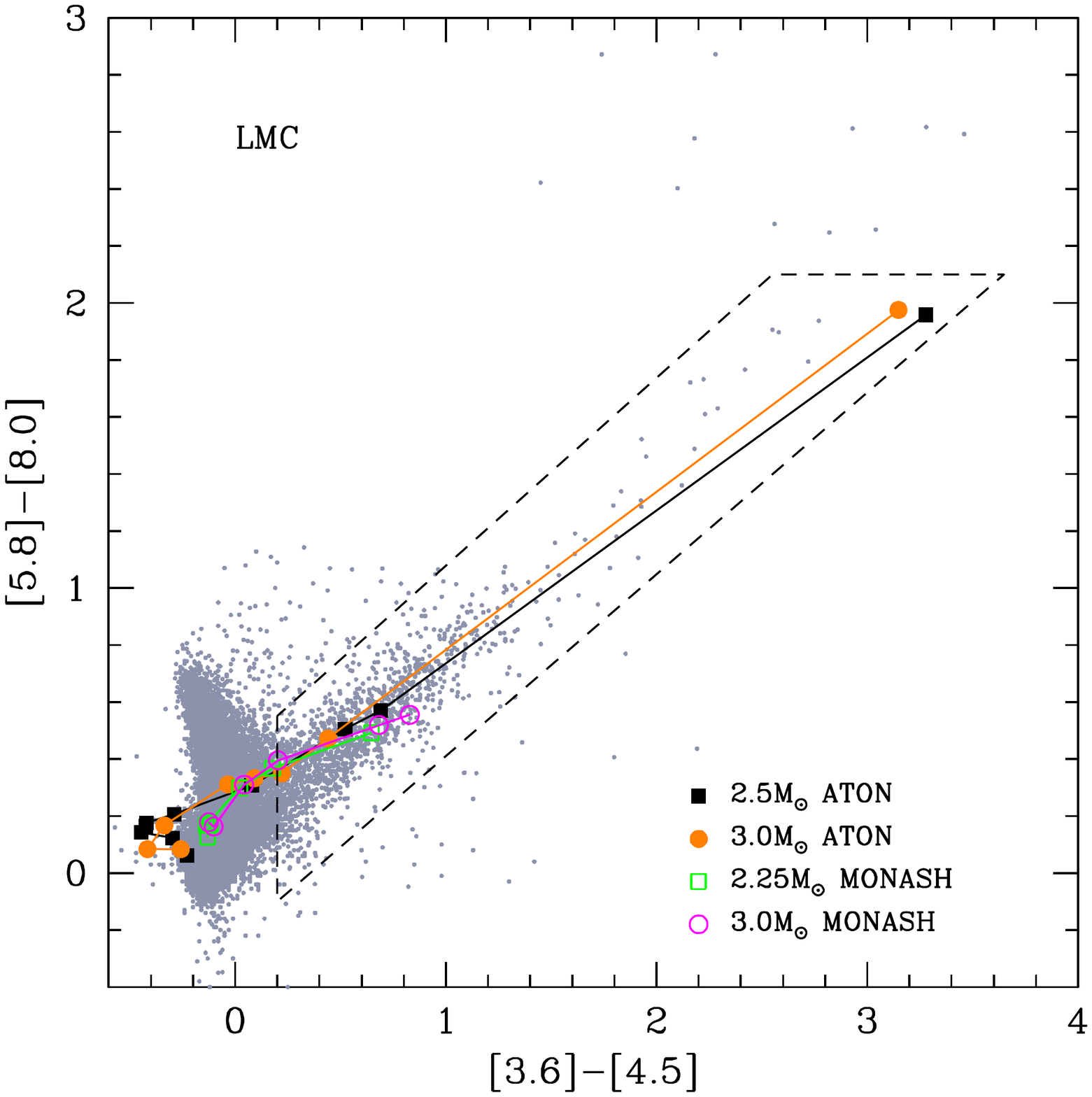}}
\end{minipage}
\vskip-40pt
\caption{Left: The observations of SMC AGB stars by \citet{boyer11} in the
colour--colour $[3.6]-[4.5],[5.8]-[8.0]$ diagram. The $1.5~M_{\odot}$ evolutionary
tracks calculated with the MONASH (open, blue triangles) and ATON (full, red triangles)
are shown. The region delimited by the dashed lines corresponds to the zone where
obscured C-stars evolve, according to the interpretation by \citet{flavia15b}.
Right: Observations of LMC, AGB stars by \citet{riebel12}; the evolutionary tracks
correspond to MONASH models of initial mass $2.25~M_{\odot}$ (open, green squares),
$3~M_{\odot}$ (open, magenta circles) and to ATON models of initial mass
$2.5~M_{\odot}$ (full, black squares), $3~M_{\odot}$ (full, orange circles).
The region delimited by dashed lines, similarly to the left panel, indicates
the zone populated by C-stars with an intense infrared emission \citep{flavia15a}.
}
\label{fmcs}
\end{figure*}

\section{Spitzer colours of carbon stars in the MCs}
\label{colours}
As discussed in the previous sections, the evolution of AGB stars with progenitors of
mass below $\sim 3~M_{\odot}$ is driven by the progressive enrichment of carbon in the
surface regions. The SED of these stars gradually shifts to the IR, with the peak of the 
emission moving to longer wavelengths. This is illustrated in Fig.~\ref{fsed}, where we 
show the SED of the $1.5~M_{\odot}$, $Z=4\times 10^{-3}$ (left panel) and $2.5~M_{\odot}$, 
$Z=8\times 10^{-3}$ models, in four evolutionary phases from the beginning of the 
AGB evolution, with practically no dust effects, to the very final phases. 
At the end of the AGB phase the SED of the $2.5~M_{\odot}$ model peaks at longer 
wavelengths compared to the $1.5~M_{\odot}$ case, 
for the reasons given in Section \ref{dustgrain}.

Fig.~\ref{fmcs} shows AGB stars in the SMC (left panel) and LMC (right) on the 
colour--colour ($[3.6]-[4.5], [5.8]-[8.0]$) plane. The regions delimited by the 
dashed lines are the C-star locii, according to
\citet{flavia15a, flavia15b}. As discussed in section \ref{mcs}, most of the obscured 
C-stars in the LMC descend from $\sim 2.5-3~M_{\odot}$ progenitors, whereas in the SMC the 
objects with the reddest IR colours are the progeny of $\sim 1.5~M_{\odot}$ stars. 
Therefore the SEDs, corresponding to the end of the AGB phase (dashed lines),
shown in Fig.~\ref{fsed}, provide an indication of the typical colours of the most
obscured AGB stars in the two galaxies. The synthetic SEDs confirm that the most
obscured LMC AGB stars reach redder $[3.6]-[4.5]$ colours than their SMC
counterparts. This explains why the C-star locus in the LMC is more
extended towards the red (up to $[3.6]-[4.5] \sim 3$) than in the SMC
($[3.6]-[4.5] \sim 1.2$)

This explanation is in constrast to the interpretation by \citet{jacco08},
who invoked a metallicity dependence on dust formation to explain the stronger dust emission 
shown by carbon stars in the LMC compared to the SMC. Note that the present
working hypothesis is focused on stars with the reddest colours, thus it is not
in contrast with the assumption that C-star formation is easier in the SMC becaue of 
the lower average metallicity \citep{sloan08}.

The overall population of C-stars in the MCs is composed of a range of masses and 
chemistries \citep{flavia15a, flavia15b}, which in the general case prevents us
from identifying the progenitors of the 
individual sources observed. To overcome this difficulty we focus on the AGB stars
populating the terminal side of the obscuration sequences of C-stars. 
As discussed in section \ref{mcs}, each star becoming C rich spends only a very
small percentage of the AGB in the last evolutionary phase with the highest
degree of obscuration. Therefore only stars of masses and metallicities given above can be
observed in such advanced evolutionary phases.

Fig.~\ref{fmcs} shows the evolutionary tracks corresponding
to the same MONASH and ATON models shown in Fig.~\ref{ftau}.
All the evolutionary sequences reproduce the diagonal band where C-stars evolve.
The only discrepancy between models and the observations concerns the C-stars
populating the bluest region of the CCD, centred at 
$([3.6]-[4.5], [5.8]-[8.0]) = (-0.1, 0.5)$. This group of stars present no signature of
dust, consequently the synthetic spectrum is entirely determined by the GRAMS SED 
used to model the atmosphere of the central star. 
The reason for the difference is the absorption feature due to CO and
C$_{3}$ molecules, which is not included in the GRAMS models. This is however
not an issue for the present analysis, because this feature 
vanishes when dust is present in the circumstellar envelope \citep{boyer11}.

The fact that both MONASH and ATON sequences trace the observations further reinforces 
the main conclusions by \citet{flavia15a, flavia15b}: the diagonal bands in 
the CCD of the LMC and SMC are obscuration sequences. The stars with the reddest colours 
correspond to those more enriched in carbon, with a larger amount of carbonaceous 
dust in the circumstellar envelope.

Fig.~\ref{fmcs} shows that the ATON models nicely reproduce the observations, the 
evolutionary tracks reaching the terminal points of the observed sequences, both in the SMC
and LMC cases. The MONASH models reproduce most of the AGB stars along the obscuration
sequences, yet they do not extend to the reddest IR colours observed.
This can be explained on the basis of the discussion in the previous section and of the 
results shown in Fig.~\ref{ftau} and \ref{fstrat}. For the SMC, the reason for the higher 
degree of obscuration reached by the ATON sequences is mainly related to the higher quantity of
carbon accumulated in $1.5~M_{\odot}$ stars during the AGB. For the LMC case, 
the main reason for the discrepancy among MONASH and ATON models is the difference in the 
description of the low-temperature molecular opacities in C-rich environments and in the 
mass-loss modelling (see section \ref{inputs}). These in turn affect the IR properties of 
the reddest AGB stars in the LMC, which reflects the mass-loss rate experienced in the final 
AGB phases, which, in turn is related to the effective temperature.

These arguments lead us to conclude that reproducing the IR colours of the 
most obscured AGB stars in the SMC requires models of stars of initial mass 
$1.5~M_{\odot}$ to achieve a surface carbon abundance $X(C) > 0.005$. 
In order to reproduce the IR colours of the LMC AGB stars, we conclude that the main
physical ingredient is the low-temperature opacities used in the calculations.
The use of low-temperature opacities that follow the chemical composition of the
envelope is especially important for stars with masses $2.5-3~M_{\odot}$, which in the last 
phases reach surface carbon abundances $X(C) > 0.01$. Furthermore, the description of mass 
loss in C-rich environments, where significant amounts of dust form, must be 
based on hydrodynamical wind models including dust formation in radiative transfer
equations.

\begin{table*}
\begin{center}
\caption{} 
\begin{tabular}{c|c|c|c|c|c|c|c|c|c|c|}
\hline
\hline 
 & $[3.6]-[4.5]$ & $\tau_{10}$ & $a_{SiC} (\mu m)$ & $a_C(\mu m)$ & Age (Gyr) & 
 $M/M_{\odot}$ & $L/(10^3L_{\odot})$ & $T_{\rm eff}$ & $\dot M/(10^{-5}M_{\odot}/$yr) & Z \\  
\hline
LMC & 3.4 & 4.5 & 0.082 & 0.28 & 0.4--0.7 &  2.5--3 & $8-10$ & 2000 & 15 & $8\times 10^{-3}$ \\
SMC & 1.2 & 0.7 & 0.063 & 0.21 & 1.5      &  1.5    & $6-7$  & 2500 & 3  & $4\times 10^{-3}$ \\
\hline 
\hline 
\end{tabular}
\end{center}
\label{tabrates}
\end{table*}

We stress that the results obtained by MONASH and ATON codes are fairly similar 
for most of the AGB phase, for the masses and metallicities investigated here. Significant 
differences are found only during the two last interpulse phases. For this reason,
the change in the physical ingredients discussed are crucial for a correct
determination of the dust properties of these stars and, more generally, for the interpretation
of IR properties of more complex stellar populations, where a significant population
of C-stars is expected. On the other hand, we expect little effects for the other
results not directly related to dust, primarily the stellar yields, which reflect the
whole AGB phase. 

\section{What do we learn from obscured carbon stars in the MCs?}
The models presented here nicely reproduce the mid--infrared colours for the SMC 
and LMC. According to our schematization AGB stars become more and more obscured as the 
carbon in the envelope increases. If this is correct, the LMC proves the ideal environment 
to host C-stars with an extremely intense infrared emission. This is because the SFH of the LMC 
peaks at $\sim 500$ Myr \citep{harris09}, the formation epoch of $\sim 2.5-3~M_{\odot}$ 
stars which are those reaching the largest abundances of carbon in the final evolutionary 
phases (see Fig.~\ref{fmodel}). If this understanding is confirmed, we reach the general 
conclusion that the degree of obscuration of the reddest stars in the LMC is the highest 
among C-rich AGB stars; for stars exhibiting a higher degree of obscuration, 
an AGB origin can be disregarded.

Table 1 reports the main properties of the individual stars which according to our interpretation
populate the reddest portion of the C-star sequence in the CCD of the MCs. Observational 
confirmation of these quantities (in particular of the predicted radial stratification of the
SiC and carbon grains) would be ideal to assess the robustness of our
modelling and interpretation. Unfortunately this is challenging with
present astronomical instrumentation. 

Concerning the synthetic SEDs of the stars examined here, we note that the dashed track 
in the right panel of Fig.~\ref{fsed} is extremely similar to the SED of the extreme carbon 
star shown in the top-left panel of Figure 7 by Riebel et al. (2012), which
is taken as a representative member of the group of most obscured C-stars in the LMC.
Interestingly, in SMC C-rich AGB stars the C$_{2}$H$_{2}$ and SiC features are
observed to be stronger and weaker, respectively, than in the LMC stars (e.g.,
Lagadec et al. 2007; Zijlstra et al. 2006, van Loon et al. 2006, 2008). 
Indeed, our synthetic SEDs, which show an
intermediate stage after becoming C-rich (red-dotted lines in Fig.~\ref{fsed}),
reproduce well this observation. The 3.1 $\mu$m C$_{2}$H$_{2}$ $+$ HCN
absorption feature is stronger in a $1.5~M_{\odot}$ SMC AGB star than in the $3~M_{\odot}$
LMC case, while the opposite is seen for the $\sim$11.5 $\mu$m SiC feature. This is 
in agreement with our framework because the production of SiC (see discussion in section 
\ref{dustmodel}) scales with the amount of silicon available at the surface, which in turn, 
is proportional to the metallicity of the stars. Therefore production of SiC is favoured 
in the LMC, because the metallicity of the C-stars in this galaxy, particularly those 
forming the largest quantities of dust, is generally larger than their counterparts in the 
SMC. On the other hand carbon dust production is approximately independent of
metallicity, because the carbon accumulated to the surface of these stars is of 
primary origin.

As for the global properties of these objects, \citet{riebel12} find that the luminosities
of the most obscured C-stars in the LMC are within the range 
$6\times 10^3 < L/L_{\odot} < 10^4$, which nicely overlaps with the range of
luminosities for the LMC reported in Table 1. 

Concerning the mass loss rates, \citet{riebel12} find that the C-stars in the LMC with
the highest degree of obscuration inject into the interstellar medium carbonaceous 
dust with a rate of $\sim 10^{-7} M_{\odot}/$yr. This is translated into an overall
mass-loss rate of $\sim 2\times 10^{-5} M_{\odot}/$yr\footnote{To find the mass-loss
rate from the dust rate we adopted a gas/dust ratio $\sim 200$. This is indeed what we
find based on our models, with a carbon mass fraction slightly above $1\%$ and a
percentage of gaseous carbon condensed into dust of $\sim 50\%$.}. This quantity is
significantly smaller than the result reported in Table 1, which is in much better
agreement with the results published by \citet{vista, vanloon06, martin07}.

On the chemical side, we have seen that the conclusions we may draw for the SMC
and LMC are different. In the former case, we showed that within the context of
our description of the AGB and dust systems the colours of the C-stars exhibiting
the highest degree of obscuration are reproduced only if the amount of carbon
accumulated at the surface exceeds $\sim 0.5\%$ of the total mass. This
C abundance this corresponds to a C/O ratio above 3. The IR colours of the most 
obscured LMC C-stars are however less sensitive to the surface carbon abundance 
than their SMC counterparts. The uncertainties affecting the degree of carbon
enrichment are still significant, which is confirmed
by the differences between the MONASH (surface carbon mass fraction slightly
below 0.02, C/O$\sim$10) and ATON (X(C)$\sim$0.01, C/O$\sim$3) models. 

The most obscured C-stars in the MCs are very faint in the optical domain but
much brighter in the near-IR. Thus the best observational way to test the
different nature (i.e., progenitor masses) of the obscured C-stars in the SMC
and LMC would be to obtain follow-up high-resolution 
($R={\lambda\over \delta \lambda}>$20,000) near-IR (JHK
bands) spectroscopy of these stars. This can be done with actual instrumentation
(e.g., CRIRES at ESO/VLT). From such spectra one could get the CNO elemental
abundances, which are expected to be different in both types of stars (see e.g.,
Ventura et al. 2015b).

On the other hand, an analysis of the chemical composition of PNe in the MCs
would provide important complementary information on the chemical enrichment
of C-stars in these galaxies. This is because only small surface abundance changes
are expected from the final AGB phase to the PN stage. For example, \citet{ventura15b} 
showed that the most C-enriched PNe in the LMC display carbon abundances below
X(C)$\sim$0.01, in agreement with the $2.5~M_{\odot}$ and $3~M_{\odot}$ ATON models and
with the $2.25~M_{\odot}$ MONASH model (see the bottom-right panel in Fig.~\ref{fmodel}). 
The extension of this type of analysis to a much wider sample of PNe in the LMC (and 
also to the SMC PNe) will help us to understand whether this upper limit in the carbon
abundance is real, or due to the limited sample of PNe examined.

\section{Conclusions}
\label{concl}
We use {\it Spitzer} IR data of C-stars in the Magellanic Clouds to improve our understanding of 
the evolution of AGB stars with masses below  $\sim 3~M_{\odot}$, which become carbon
stars through repeated TDU events. We focus on the objects with the highest degree
of obscuration,  because it is easier to characterise their progenitors in terms of their 
initial masses and metallicities. This is due to differences in the SFH between the 
SMC and the LMC, allowing us to investigate stars of different mass. The analysis of the 
SMC permits to study the evolutionary properties of AGB stars of mass $\sim 1.5~M_{\odot}$, 
whereas we use the LMC to investigate higher mass stars, with $\sim 2.5-3~M_{\odot}$. 

The LMC sequence of C-stars extends to redder IR colours ($[3.6]-[4.5] \sim 3$) compared 
to the SMC ($[3.6]-[4.5] \sim 1.2$). According to our interpretation, this is because of 
the higher quantity of carbonaceous dust forming in the circumstellar envelope 
of $\sim 2.5-3~M_{\odot}$ stars, in comparison with their lower mass ($\sim 1.5~M_{\odot}$)
counterparts evolving in the SMC.

The comparison between models and observations suggests the following: a) in 
$\sim 1.5~M_{\odot}$ AGB stars TDU must be efficient enough to produce a surface carbon 
abundance above $X(C) \sim 0.005$ before the end of the AGB phase; b) to correctly 
model the late AGB phases of stars of mass $\sim 2.5-3~M_{\odot}$ it is mandatory to 
include the carbon enrichment in the calculation of the low-temperature molecular 
opacities \citep{marigo02,vm09,vm10,kamath12}. In addition, the description of mass loss 
must rely on hydrodynamical simulations, which take into account dust formation. These 
physical ingredients are crucial in order to reproduce the extremely red colours of the 
C-stars in the LMC with the largest degree of obscuration. Based on our modelling,
these stars evolve at effective temperatures close to $\sim 2000$ K and loose their
external mantle with a rate slightly above $10^{-4}~M_{\odot}/$yr.

If our understanding is correct, we expect that the most obscured stars in the SMC 
are surrounded by a more internal region with SiC dust grains
$\sim 0.06 \mu$m sized, and a more external zone, $\sim 10$ stellar radii from the 
surface of the star, where SiC grains and $\sim 0.2~\mu$m sized, solid carbon 
particles are present. In the counterparts in the LMC, the higher metallicity favours a larger
growth of SiC grains, which reach dimensions of the order of $\sim 0.06~\mu$m.
Solid carbon grains on the other hand can grow to $\sim 0.28~\mu$m, because
these stars are more enriched in gaseous carbon than their counterparts 
of smaller mass.

High-resolution near-IR spectroscopy of the most obscured C-stars in the SMC and LMC as 
well as the analysis of a wider MC PNe sample will allow us to test the conclusions reached 
here; i.e., the most obscured C-rich AGB stars in the LMC are the descendants of stars 
with higher initial masses than their SMC counterparts.

\section*{Acknowledgments}
The authors are indebted to the referee, J. Th. van Loon, for the careful reading of the
manuscript and for the several comments, that help improving significantly the quality
of this work. P.V. was supported by PRIN MIUR 2011 "The Chemical and Dynamical Evolution 
of the Milky Way and Local Group Galaxies" (PI: F. Matteucci), prot. 2010LY5N2T. A.I.K. 
was supported through an Australian Research Council Future Fellowship (FT110100475). 
DAGH was funded by the Ram\'on y Cajal fellowship number RYC$-$2013$-$14182 and he 
acknowledges support 
provided by the Spanish Ministry of Economy and Competitiveness (MINECO) under grant 
AYA$-$2014$-$58082-P. F.D. and M.D.C. acknowledge support from the Observatory of Rome.


\begin{thebibliography}{99}
\bibitem[\protect\citeauthoryear{Allard et al. et al.}{2012}]{nextgen}
Allard F., Homeier D., Freytag B., 2012, Philosophical Transactions of the 
Royal Society A: Mathematical, Physical and Engineering Sciences, vol. 370, 
issue 1968, p2765
\bibitem[Becker \& Iben(1980)]{becker80} Becker S.~A., Iben I. Jr.\ 1980, ApJ, 237, 111 
\bibitem[Bessell et al.(1989)]{bessel89} Bessell M.~S., Brett J.~M., Wood P.~R., 
Scholz M.\ 1989, A\&AS, 77, 1
\bibitem[Bladh \& H{\"o}fner(2012)]{bladh12} Bladh S., H{\"o}fner S.\ 2012, A\&A, 546, 76
\bibitem[\protect\citeauthoryear{Bl\"ocker}{1995}]{blocker95}
Bl\"ocker T., 1995, A\&A, 297, 727
\bibitem[\protect\citeauthoryear{Bl\"ocker \& Sch\"onberner}{1991}]{blocker91}
Bl\"ocker T., Sch\"onberner D., 1991, A\&A, 244, L43
\bibitem[\protect\citeauthoryear{Blum et al.}{2006}]{blum06}
Blum R.~D. et al., 2006, AJ, 132, 2034
\bibitem[Boyer et al.(2011)]{boyer11} Boyer M.~L., Srinivasan S., 
van Loon, J.~T., et al.\ 2011, AJ, 142, 103 
\bibitem[\protect\citeauthoryear{Boyer et al.}{2012}]{boyer12}
Boyer M.~L., Srinivasan S., Riebel D., McDonald I., van Loon J.~Th., Clayton G.~C.,
Gordon K.~D., Meixner M., Sargent B.~A., Sloan G.~C., 2012, ApJ, 748, 40
\bibitem[\protect\citeauthoryear{Boyer et al.}{2015}]{boyer15}
Boyer M.~L., McDonald I., Srinivasan S., et al. 2015, ApJ, 810, 116
\bibitem[\protect\citeauthoryear{Canuto \& Mazzitelli}{1991}]{cm91}
Canuto V.~M.~C., Mazzitelli I., 1991, ApJ, 370, 295
\bibitem[\protect\citeauthoryear{Carrera et al.}{2008}]{carrera08}
Carrera R., Gallart C., Hardy E., Aparicio A., Zinn R., 2008, AJ, 135, 836
\bibitem[Chiosi et al.(1993)]{chiosi93} Chiosi C., Wood P.~R., Capitanio N.\ 1993, ApJS, 
86, 541 
\bibitem[\protect\citeauthoryear{Cioni et al.}{2000a}]{cioni00a}
Cioni, M.-R. L., Loup, C., Habing, H. J., et al., 2000a, A\&AS, 144, 235
\bibitem[\protect\citeauthoryear{Cioni et al.}{2000b}]{cioni00b}
Cioni M.-R. L., Habing H. J.,  Israel F. P., 2000b, A\&A, 358, L9
\bibitem[Cioni et al.(2006)]{cioni06} Cioni M.-R.~L., Girardi L., Marigo P., 
Habing H.~J.\ 2006, A\&A, 448, 77
\bibitem[D'Antona \& Mazzitelli(1996)]{dantona96} D'Antona,F., Mazzitelli I.\ 1996, 
ApJ, 470, 1093 
\bibitem[\protect\citeauthoryear{Dell'Agli et al.}{2014}]{flavia14}
Dell'Agli F., Ventura P., Garc{\'{\i}}a-Hern{\'a}ndez D.~A., Schneider R., Di Criscienzo M., 
Brocato E., D'Antona F., Rossi C., 2014, MNRAS, 442, L38
\bibitem[\protect\citeauthoryear{Dell'Agli et al.}{2015a}]{flavia15a}
Dell'Agli F., Ventura P., Schneider R., Di Criscienzo M., Garc{\'{\i}}a-Hern{\'a}ndez D.~A.,  
Rossi C., Brocato E.  2015a, MNRAS, 447, 2992
\bibitem[\protect\citeauthoryear{Dell'Agli et al.}{2015b}]{flavia15b}
Dell'Agli F., Garc{\'{\i}}a-Hern{\'a}ndez D.~A., Ventura P., Schneider R., Di Criscienzo M.,   
Rossi C. 2015b, MNRAS, in press
\bibitem[\protect\citeauthoryear{Di Criscienzo et al.}{2013}]{paperIII}
Di Criscienzo M., Dell'Agli F., Ventura P., Schneider R., Valiante R., 
La Franca F., Rossi C., Gallerani S., Maiolino, R., 2013, MNRAS, 433, 313
\bibitem[Doherty et al.(2014a)]{doherty14a} Doherty C.~L., Gil-Pons P., Lau H.~H.~B., 
Lattanzio J.~C., Siess L.\ 2014a, MNRAS, 437, 195 
\bibitem[Doherty et al.(2014b)]{doherty14b} Doherty C.~L., Gil-Pons P., Lau H.~H.~B., 
Lattanzio J.~C., Siess L., Campbell S.~W. \ 2014b, MNRAS, 441, 582 
\bibitem[\protect\citeauthoryear{Epchtein et al.}{1994}]{epchtein94}
Epchtein N. et al., 1994, Ap\&SS, 217, 3
\bibitem[\protect\citeauthoryear{Feast}{1999}]{feast99}
Feast M., 1999, PASP, 111, 775
\bibitem[\protect\citeauthoryear{Ferrarotti \& Gail}{2001}]{fg01}
Ferrarotti A.~D., Gail H.~P., 2001, A\&A, 371, 133
\bibitem[\protect\citeauthoryear{Ferrarotti \& Gail}{2002}]{fg02}
Ferrarotti A.~D., Gail H.~P., 2002, A\&A, 382, 256
\bibitem[\protect\citeauthoryear{Ferrarotti \& Gail}{2006}]{fg06}
Ferrarotti A.~D., Gail H.~P., 2006, A\&A, 553, 576
\bibitem[Frost \& Lattanzio(1996)]{frost96} Frost C.~A., Lattanzio J.~C.\ 1996, ApJ, 473, 383 
\bibitem[\protect\citeauthoryear{Gail \& Sedlmayr}{1985}]{gs85}
Gail H.~P., Sedlmayr E., 1985, A\&A, 148, 183
\bibitem[\protect\citeauthoryear{Gail \& Sedlmayr}{1999}]{gs99}
Gail H.~P., Sedlmayr E., 1999, A\&A, 347, 594
\bibitem[\protect\citeauthoryear{Gordon et al.}{2011}]{gordon11}
Gordon K.~D. et al. 2011, AJ, 142, 102
\bibitem[\protect\citeauthoryear{Grevesse \& Sauval}{1998}]{gs98}
Grevesse N., Sauval A.~J, 1998, SSrv, 85, 161
\bibitem[Groenewegen \& de Jong(1993)]{martin93} Groenewegen M.~A.~T., de Jong T.\ 1993, 
A\&A, 267, 410
\bibitem[Groenewegen et al.(2007)]{martin07} 
Groenewegen M.~A.~T., Wood P.~R., Sloan G.~C., et al.\ 2007, MNRAS, 376, 313 
\bibitem[Gullieuszik et al.(2012)]{vista} Gullieuszik M., Groenewegen M.~A.~T., 
Cioni M.-R.~L., et al.\ 2012, A\&A, 537, A105
\bibitem[Harris \& Zaritsky(2004)]{harris04} Harris J., Zaritsky D.\ 2004, AJ, 127, 1531 
\bibitem[\protect\citeauthoryear{Harris \& Zaritsky}{2009}]{harris09}
Harris J., Zaritsky D. 2009, ApJ, 138, 1243
\bibitem[Herwig(2005)]{herwig05} Herwig F.\ 2005, ARA\&A, 43, 435 
\bibitem[\protect\citeauthoryear{H{\"o}fner}{2008}]{hofner08}
H{\"o}fner S. 2008, A\&A, 491, L1
\bibitem[\protect\citeauthoryear{Herwig}{2005}]{herwig05}
Herwig F., 2005, AR\&A, 43, 435
\bibitem[Iben(1982)]{iben82} Iben I. Jr.\ 1982, ApJ, 260, 821
\bibitem[Iben \& Renzini(1983)]{iben83} Iben I. Jr., Renzini A.\ 1983, ARA\&A 21, 271 
\bibitem[Izzard et al.(2004)]{izzard04} Izzard R.~G., Tout C.~A., Karakas A.~I., 
Pols O.~R.\ 2004, MNRAS, 350, 407
\bibitem[Lagadec et al.(2007)]{lagadec07} Lagadec E., Zijlstra A.~A., Sloan G.~C. et al. 
2007, MNRAS, 376, 1270
\bibitem[Lattanzio(1986)]{lattanzio86} Lattanzio J.~C.\ 1986, ApJ, 311, 708 
\bibitem[Kamath et al.(2012)]{kamath12} Kamath D., Karakas A.~I., Wood, P.~R.\ 2012, ApJ, 
746, 20 
\bibitem[Karakas et al.(2002)]{karakas02} Karakas A.~I., 
Lattanzio J.~C., Pols O.~R.\ 2002, PASA, 19, 515
\bibitem[Karakas \& Lattanzio(2007)]{karakas07} Karakas A., Lattanzio J.~C.\ 2007, PASA, 24, 103 
\bibitem[Karakas(2010)]{karakas10} Karakas A.~I.\ 2010, MNRAS, 403, 1413 
\bibitem[Karakas et al.(2010b)]{karakas10b} Karakas A.~I., Campbell S.~W., 
Stancliffe R.~J.\ 2010, ApJ, 713, 374
\bibitem[Karakas(2011)]{karakas11} Karakas A.~I.\ 2011, in: Why Galaxies Care about AGB 
Stars II: Shining Examples and Common Inhabitants, ASPC, 445, 3
\bibitem[Karakas \& Lattanzio(2014)]{karakas14} Karakas A.~I., Lattanzio J.~C.\ 2014, PASA, 31, e030 
\bibitem[Maiolino et al.(2004)]{maiolino04} Maiolino R., Schneider R., Oliva E., 
et al.\ 2004, Nature, 431, 533 
\bibitem[Marigo(2002)]{marigo02} Marigo P.\ 2002, A\&A, 387, 507 
\bibitem[\protect\citeauthoryear{Marigo \& Aringer}{2009}]{marigo09} 
Marigo P., Aringer B., 2009, A\&A, 508, 1538
\bibitem[Marigo \& Girardi(2007)]{marigo07} Marigo P., Girardi L.\ 2007, A\&A, 469, 23
\bibitem[Marigo et al.(1999)]{marigo99} Marigo P., Girardi L., Bressan A.\ 1999, A\&A, 344, 123
\bibitem[Marigo et al.(2008)]{marigo08} Marigo P., Girardi L., Bressan A., 
Groenewegen M. A. T., Silva L., Granato, G. L.\ 2008, A\&A, 482, 883
\bibitem[\protect\citeauthoryear{Matsuura et al.}{2009}]{matsuura09} 
Matsuura M., et al., 2009, MNRAS, 396, 918
\bibitem[\protect\citeauthoryear{Matsuura et al.}{2013}]{matsuura13} 
Matsuura M., Woods P.~V., Owen P.~J., 2013, MNRAS, 429, 2527
\bibitem[\protect\citeauthoryear{Meixner et al.}{2006}]{meixner06} 
Meixner M. et al. 2006, AJ, 132, 2268
\bibitem[\protect\citeauthoryear{Meixner et al.}{2010}]{meixner10} 
Meixner M. et al. 2010, A\&A, 518, L71
\bibitem[\protect\citeauthoryear{Meixner et al.}{2013}]{meixner13} 
Meixner M. et al. 2013, ApJ, 146, 62
\bibitem[\protect\citeauthoryear{Nanni et al.}{2013a}]{nanni13a} 
Nanni A., Bressan A., Marigo P., Girardi L., 2013a, MNRAS, 434, 488
\bibitem[\protect\citeauthoryear{Nanni et al.}{2013b}]{nanni13b} 
Nanni A., Bressan A., Marigo P., Girardi L., 2013b, MNRAS, 434, 2390
\bibitem[\protect\citeauthoryear{Nanni et al.}{2014}]{nanni14} 
Nanni A. Bressan A. Marigo P. Girardi L., 2014, MNRAS, 438, 2328
\bibitem[\protect\citeauthoryear{Nenkova et al.}{1999}]{dusty} 
Nenkova M., Ivezi{\'c} {\v Z}., Elitzur M., 1999, in: LPIContributions 969,
Workshop on Thermal Emission Spectroscopy and Analysis of Dust, Disks, and 
Regoliths, ed. A. Sprague, D. K. Lynch, \& M. Sitko (Houston, TX: Lunar and 
Planetary Institute), 20
\bibitem[\protect\citeauthoryear{Nozawa et al.}{2003}]{nozawa03}
Nozawa T., Kozasa T., Umeda H., Maeda K., Nomoto K. 2003, ApJ, 598, 785
\bibitem[Paczy{\'n}ski(1970)]{paczynski} Paczy{\'n}ski, B.\ 1970, 
Acta Astr., 20, 47 
\bibitem[Piatti(2012)]{piatti12} Piatti A.~E.\ 2012, MNRAS
422, 1109 
\bibitem[\protect\citeauthoryear{Piatti \& Geisler}{2013}]{piatti13} 
Piatti A.E., Geisler G., 2013, AJ, 145, 17
\bibitem[\protect\citeauthoryear{Renzini \& Voli}{1981}]{renzini81} Renzini A.,
Voli M., 1981, A\&A, 94, 175
\bibitem[\protect\citeauthoryear{Riebel et al.}{2010}]{riebel10} 
Riebel D., Meixner M., Fraser O., Srinivasan S., Cook K., Vijh U., 2010, ApJ, 723, 1195
\bibitem[\protect\citeauthoryear{Riebel et al.}{2012}]{riebel12} 
Riebel D., Srinivasan S., Sargent B., Meixner M., 2012, AJ, 753, 71
\bibitem[Romano et al.(2010)]{romano10} Romano D., Karakas A.~I., Tosi M., 
Matteucci F.\ 2010, A\&A 522, AA32 
\bibitem[\protect\citeauthoryear{Schwarzschild \& Harm}{1965}]{schw65}
Schwarzschild M.,  Harm R. 1965, ApJ, 142, 855
\bibitem[\protect\citeauthoryear{Schwarzschild \& Harm}{1967}]{schw67}
Schwarzschild M.,  Harm R. 1967, ApJ, 145, 496
\bibitem[\protect\citeauthoryear{Skrutskie et al.}{2006}]{skrutskie06} 
Skrutskie M.~F. et al., 2006, AJ, 131, 1163
\bibitem[\protect\citeauthoryear{Schlegel et al.}{1998}]{schlegel98} 
Schlegel D.~J., Finkbeiner D.~P., Davis M., 1998, ApJ, 500, 525
\bibitem[Schneider et al.(2014)]{schneider14} Schneider R., 
Valiante R., Ventura P., Dell'Agli F., Di Criscienzo M., Hirashita H.,
Kemper F. \ 2014, MNRAS, 442, 1440 
\bibitem[\protect\citeauthoryear{Skrutskie}{1998}]{skrutskie98} 
Skrutskie M. 1998, in: The Impact of Near-Infrared Sky Surveys
on Galactic and Extragalactic Astronomy, Proc. of the 3rd Euroconference 
on Near-Infrared Surveys, ed. N. Epchtein, Astrophysics and Space Science 
library (Dordrecht: Kluwer), 230, 11 
\bibitem[Sloan et al.(2008)]{sloan08} Sloan G. C., Kraemer K. E., Wood P. R., 
Zijlstra A. A., Bernard-Salas J., Devost D., Houck, J. R. \ 2008, ApJ, 686, 1056 
\bibitem[\protect\citeauthoryear{Srinivasan et al.}{2009}]{srinivasan09} 
Srinivasan S. et al., 2009, AJ, 137, 4810
\bibitem[Srinivasan et al.(2011)]{srinivasan11} Srinivasan S., Sargent B.~A., 
Meixner, M.\ 2011, A\&A, 532, A54 
\bibitem[Stancliffe et al.(2005)]{stancliffe05} Stancliffe R.~J., Izzard R.~G.,
Tout C.~A.\ 2005, MNRAS, 356, L1 
\bibitem[Todini \& Ferrara (2001)]{todini01} Todini P., Ferrara A. 2001, MNRAS, 325, 726
\bibitem[\protect\citeauthoryear{Valiante et al.}{2009}]{valiante09}
Valiante R., Schneider R., Bianchi S., Andersen A., Anja C., 2009, MNRAS, 397, 1661
\bibitem[\protect\citeauthoryear{van Loon et al.}{1999}]{jacco99} van Loon J.~Th., 
Zijlstra A.~A., Groenewegen M.~A.~T. 1999, A\&A, 346, 805
\bibitem[van Loon(2006)]{vanloon06} van Loon, J.\ 2006, 
Astronomical Society of the Pacific Conference Series, 357, 155 
\bibitem[van Loon et al.(2006)]{jacco06} van Loon J. Th., Marshall J. R., Cohen M., 
Matsuura M., Wood P. R., Yamamura I., Zijlstra A. A. \ 2006, A\&A, 447, 971 
\bibitem[van Loon et al.(2008)]{jacco08} van Loon J.~T., Cohen M., Oliveira J.~M., 
 Matsuura M., McDonald I., Sloan G.~C., Wood P.~R., Zijlstra A.~A. \ 2008, A\&A, 487, 1055
\bibitem[Vassiliadis \& Wood(1993)]{VW93} Vassiliadis E., Wood P.~R.\ 1993, ApJ, 413, 641 
\bibitem[Ventura et al.(2001)]{ventura01} Ventura P., D'Antona F., Mazzitelli I., 
Gratton R.\ 2001, ApJ, 550, L65 
\bibitem[\protect\citeauthoryear{Ventura \& D'Antona}{2005}]{vd05}
Ventura P., D'Antona F., 2005, A\&A, 431, 279
\bibitem[\protect\citeauthoryear{Ventura \& D'Antona}{2008}]{ventura08} 
Ventura P., D'Antona F., 2008, A\&A, 479, 805
\bibitem[\protect\citeauthoryear{Ventura \& D'Antona}{2009}]{ventura09} 
Ventura P., D'Antona F., 2009, MNRAS, 499, 835
\bibitem[Ventura et al.(2001)]{ventura01} Ventura P., D'Antona F., Mazzitelli I., 
Gratton R.\ 2001, ApJL, 550, L65 
\bibitem[\protect\citeauthoryear{Ventura et al.}{2012a}]{paperI} 
Ventura P., Di Criscienzo M., Schneider R., Carini R., Valiante R., D'Antona F., 
Gallerani S., Maiolino R., Tornamb\'e A., 2012a, MNRAS, 420, 1442
\bibitem[\protect\citeauthoryear{Ventura et al.}{2012b}]{paperII} 
Ventura P., Di Criscienzo M., Schneider R., Carini R., Valiante R., D'Antona F., 
Gallerani S., Maiolino R., Tornamb\'e A., 2012b, MNRAS, 424, 2345
\bibitem[\protect\citeauthoryear{Ventura et al.}{2014a}]{paperIV} 
Ventura P., Dell'Agli F., Di Criscienzo M., Schneider R., Rossi C., La Franca F., 
Gallerani S., Valiante R., 2014a, MNRAS, 439, 977
\bibitem[\protect\citeauthoryear{Ventura et al.}{2013}]{ventura13} 
Ventura P., Di Criscienzo, M., Carini R., D'Antona F., 2013, MNRAS, 431, 3642
\bibitem[Ventura et al.(2015a)]{ventura15a} Ventura P., Karakas A.~I., Dell'Agli F., 
Boyer M.~L., Garc{\'{\i}}a-Hern{\'a}ndez D.~A., Di Criscienzo M., Schneider R. 2015a, 
MNRAS, 450, 3181 
\bibitem[\protect\citeauthoryear{Ventura \& Marigo}{2009}]{vm09} Ventura P.,
Marigo P., 2009, MNRAS, 399, L54
\bibitem[\protect\citeauthoryear{Ventura \& Marigo}{2010}]{vm10} Ventura P.,
Marigo P., 2010, MNRAS, 408, 2476
\bibitem[\protect\citeauthoryear{Ventura et al.}{2015b}]{ventura15b} Ventura P., 
Stanghellini L., Dell'Agli F., Garc{\'{\i}}a-Hern{\'a}ndez D.~A., Di Criscienzo, M.\ 2015b, 
MNRAS, 452, 3679 
\bibitem[\protect\citeauthoryear{Ventura et al.}{1998}]{ventura98} Ventura P.,
Zeppieri A., Mazzitelli I., D'Antona F., 1998, A\&A, 334, 953
\bibitem[\protect\citeauthoryear{Wachter et al.}{2002}]{wachter02} 
Wachter A., Schr\"oder K.~P., Winters J.~M., Arndt T.~U., Sedlmayr E., 2002, A\&A, 384, 452
\bibitem[\protect\citeauthoryear{Wachter et al.}{2008}]{wachter08} 
Wachter A., Winters J.~M., Schr\"oder K.~P., Sedlmayr E., 2008, A\&A, 486, 497
\bibitem[\protect\citeauthoryear{Weisz et al.}{2013}]{weisz13} 
Weisz D.~R., Dolphin A.~E., Skillman E.~D., Holtzman J., Dalcanton J.~J., 
Cole A.~A., Neary K., 2013, MNRAS, 431, 364
\bibitem[\protect\citeauthoryear{Wood}{1979}]{wood79} Wood P.~R., 2004, ApJ, 227, 220
\bibitem[Woods et al.(2011)]{woods11} Woods, P.~M., Oliveira, 
J.~M., Kemper, F., et al.\ 2011, MNRAS, 411, 1597 
\bibitem[Yang et al.(2004)]{yang04} Yang X., Chen P., He J. 2004, A\&A, 414, 1049
\bibitem[\protect\citeauthoryear{Zaritsky et al.}{2004}]{zaritsky04}
Zaritsky D., Harris J., Thompson I.~B., Grebel E.~K. 2004, AJ, 128, 1606
\bibitem[\protect\citeauthoryear{Zhukovska et al.}{2008}]{zhukovska08}
Zhukovska S., Gail H.-P., Trieloff M., 2008, A\&A, 479, 453
\bibitem[\protect\citeauthoryear{Zhukovska \& Henning}{2013}]{zhukovska13}
Zhukovska S., Henning T., 2013, A\&A, 555, 99
\bibitem[\protect\citeauthoryear{Zijlstra et al.}{2006}]{albert06} Zijlstra A.~A., 
Matsuura M., Wood P.~R. et al. 2006, MNRAS, 370, 1961
\end{thebibliography}
\end{document}